\documentclass[tighten,times,twocolumn]{aastex63}

\graphicspath{{./}{figures/}}

\usepackage{amsmath}
\usepackage{mathtools}

\renewcommand{\vec}[1]{{\boldsymbol{\mathbf{#1}}}}

\DeclareMathOperator{\sinc}{sinc}

\newcommand{\appropto}{\mathrel{\vcenter{
  \offinterlineskip\halign{\hfil$##$\cr
    \propto\cr\noalign{\kern2pt}\sim\cr\noalign{\kern-2pt}}}}}

\usepackage{hyperref}
\usepackage{xspace}
\usepackage{graphicx}
\usepackage[enable]{easy-todo}
\usepackage{hyperref}

\usepackage{savesym}
\savesymbol{tablenum}
\usepackage{siunitx}
\restoresymbol{SIX}{tablenum}

\newcommand{\tcm}[0]{{\SI{21}{cm}} }

\definecolor{NavyBlue}{HTML}{006EB8}

\begin{document}

\title{Holographic Beam Measurements of the Canadian Hydrogen Intensity Mapping Experiment (CHIME)}
\date{\today}

\newcommand{\ASIAA}{Institute of Astronomy and Astrophysics, Academia Sinica, Astronomy-Mathematics Building, No. 1, Sec. 4, Roosevelt Road, Taipei 10617, Taiwan}
\newcommand{\ASTRON}{ASTRON, Netherlands Institute for Radio Astronomy, Oude Hoogeveensedijk 4, 7991 PD Dwingeloo, The Netherlands}
\newcommand{\BF}{Banting Fellow}
\newcommand{\CALTECH}{Division of Physics, Mathematics, and Astronomy, California Institute of Technology, Pasadena, CA 91125, USA}
\newcommand{\CIFAR}{Canadian Institute for Advanced Research, MaRS Centre, West Tower, 661 University Ave, Suite 505, Toronto, ON, M5G 1M1 Canada}
\newcommand{\CIFARAZGS}{CIFAR Azrieli Global Scholars Program, MaRS Centre, West Tower, 661 University Ave, Suite 505, Toronto, ON, M5G 1M1 Canada}
\newcommand{\CITA}{Canadian Institute for Theoretical Astrophysics, 60 St.~George Street, Toronto, ON M5S 3H8, Canada}
\newcommand{\CMU}{McWilliams Center for Cosmology, Department of Physics, Carnegie Mellon University, Pittsburgh, PA 15213, USA}
\newcommand{\CSRO}{CSIRO Space & Astronomy, Parkes Observatory, P.O. Box 276, Parkes NSW 2870, Australia}
\newcommand{\CTK}{Cahill Center for Astronomy and Astrophysics, California Institute of Technology, 1216 E California Boulevard, Pasadena, CA 91125, USA}
\newcommand{\CU}{Cornell Center for Astrophysics and Planetary Science, Cornell University, Ithaca, NY 14853, USA}
\newcommand{\CUT}{Department of Space, Earth and Environment, Chalmers University of Technology, Onsala Space Observatory, 439 92, Onsala, Sweden}
\newcommand{\DAA}{David A.~Dunlap Department of Astronomy \& Astrophysics, University of Toronto, 50 St.~George Street, Toronto, ON M5S 3H4, Canada}
\newcommand{\DF}{Dunlap Fellow}
\newcommand{\DI}{Dunlap Institute for Astronomy \& Astrophysics, University of Toronto, 50 St.~George Street, Toronto, ON M5S 3H4, Canada}
\newcommand{\DRAO}{Dominion Radio Astrophysical Observatory, Herzberg Research Centre for Astronomy and Astrophysics, National Research Council Canada, PO Box 248, Penticton, BC V2A 6J9, Canada}
\newcommand{\FRQNT}{FRQNT Postdoctoral Fellow}
\newcommand{\ICRARC}{International Centre for Radio Astronomy Research, Curtin University, Bentley, WA 6102, Australia}
\newcommand{\IU}{Istanbul University, Science Faculty, Department of Astronomy and Space Sciences, Beyaz\i t, 34119, \.Istanbul, T\"urkiye}
\newcommand{\IUOBS}{
Istanbul University Observatory Research and Application Center, Istanbul University 34119, \.Istanbul T\"urkiye}
\newcommand{\MITK}{MIT Kavli Institute for Astrophysics and Space Research, Massachusetts Institute of Technology, 77 Massachusetts Ave, Cambridge, MA 02139, USA}
\newcommand{\MITP}{Department of Physics, Massachusetts Institute of Technology, 77 Massachusetts Ave, Cambridge, MA 02139, USA}
\newcommand{\MSI}{Trottier Space Institute, McGill University, 3550 rue University, Montr\'eal, QC H3A 2A7, Canada}
\newcommand{\MSIF}{McGill Space Institute Fellow}
\newcommand{\MU}{Department of Physics, McGill University, 3600 rue University, Montr\'eal, QC H3A 2T8, Canada}
\newcommand{\MPIfR}{Max-Planck-Institut f{\"u}r Radioastronomie, Auf dem H{\"u}gel 69, D-53121 Bonn, Germany}
\newcommand{\NCRA}{National Centre for Radio Astrophysics, Post Bag 3, Ganeshkhind, Pune, 411007, India}
\newcommand{\NHFP}{NHFP Einstein Fellow}
\newcommand{\NRAO}{National Radio Astronomy Observatory, 520 Edgemont Rd, Charlottesville, VA 22903, USA}
\newcommand{\NU}{Center for Interdisciplinary Exploration and Research in Astrophysics (CIERA) and Department of Physics and Astronomy, Northwestern University, Evanston, IL 60208, USA}
\newcommand{\PI}{Perimeter Institute for Theoretical Physics, 31 Caroline Street N, Waterloo, ON N25 2YL, Canada}
\newcommand{\SR}{Sidrat Research, PO Box 73527 RPO Wychwood, Toronto, ON M6C 4A7, Canada}
\newcommand{\TIFR}{Department of Astronomy and Astrophysics, Tata Institute of Fundamental Research, Mumbai, 400005, India}
\newcommand{\TSI}{Trottier Space Institute, McGill University, 3550 rue University, Montr\'eal, QC H3A 2A7, Canada}
\newcommand{\UBC}{Department of Physics and Astronomy, University of British Columbia, 6224 Agricultural Road, Vancouver, BC V6T 1Z1 Canada}
\newcommand{\UCBASTRO}{Department of Astronomy, University of California Berkeley, Berkeley, CA 94720, USA}
\newcommand{\UBCO}{Department of Computer Science, Math, Physics, \& Statistics, University of British Columbia, Okanagan Campus, Kelowna, BC V1V 1V7, Canada}
\newcommand{\UCSCa}{Division of Physical and Biological Sciences, University of California Santa Cruz, Santa Cruz, CA 95064, USA}
\newcommand{\UCSCb}{Department of Astronomy and Astrophysics, University of California Santa Cruz, Santa Cruz, CA 95064, USA}
\newcommand{\UM}{Department of Physics and Astronomy, University of Manitoba, Winnipeg, MB R3T 2N2, Canada}
\newcommand{\UTPHYS}{Department of Physics, University of Toronto, 60 St.~George Street, Toronto, ON M5S 1A7, Canada}
\newcommand{\UVA}{Anton Pannekoek Institute for Astronomy, University of Amsterdam, Science Park 904, 1098 XH Amsterdam, The Netherlands}
\newcommand{\UW}{Department of Physics and Astronomy, University of Waterloo, Waterloo, ON N2L 3G1, Canada}
\newcommand{\UWM}{Department of Physics, University of Wisconsin-Madison, 1150 University Ave, Madison, WI 53706, USA}
\newcommand{\VENI}{Veni Fellow}
\newcommand{\WVUGWAC}{Center for Gravitational Waves and Cosmology, West Virginia University, Chestnut Ridge Research Building, Morgantown, WV 26505, USA}
\newcommand{\WVULCSEE}{Lane Department of Computer Science and Electrical Engineering, 1220 Evansdale Drive, PO Box 6109, Morgantown, WV 26506, USA}
\newcommand{\WVUPA}{Department of Physics and Astronomy, West Virginia University, PO Box 6315, Morgantown, WV 26506, USA }
\newcommand{\YORK}{Department of Physics and Astronomy, York University, 4700 Keele Street, Toronto, ON MJ3 1P3, Canada}
\newcommand{\YU}{Department of Physics, Yale University, New Haven, CT 06520, USA}
\newcommand{\SFU}{Department of Statistics & Actuarial Science, Simon Fraser University, 8888 University Dr, Burnaby, BC V5A 1S6, Canada}
\newcommand{\ASU}{Department of Physics, Arizona State University, Tempe, AZ, USA}
\newcommand{\RRI}{Raman Research Institute, Bangalore, India}

\shortauthors{CHIME Collaboration}
\collaboration{100}{The CHIME Collaboration:}
\author[0000-0001-6523-9029]{Mandana Amiri}
\affiliation{\UBC}
\author[0000-0002-7758-9859]{Arnab Chakraborty}
\affiliation{\MU}
\affiliation{\TSI}
\author[0000-0002-0190-2271]{Simon Foreman}
\affiliation{\ASU}

\author[0000-0002-1760-0868]{Mark Halpern}
\affiliation{\UBC}
\author[0000-0001-7301-5666]{Alex S Hill}
\affiliation{\UBCO}
\affiliation{\DRAO}
\author[0000-0002-4241-8320]{Gary Hinshaw}
\affiliation{\UBC}
\author{T.L. Landecker}
\affiliation{\DRAO}
\author[0000-0001-8064-6116]{Joshua MacEachern}
\affiliation{\UBC}
\author[0000-0002-4279-6946]{Kiyoshi W. Masui}
\affiliation{\MITK}
\affiliation{\MITP}
\author[0000-0002-0772-9326]{Juan Mena-Parra}
\affiliation{\DI}
\affiliation{\DAA}
\author[0000-0001-8292-0051]{Nikola  Milutinovic}
\affiliation{\UBC}
\author[0000-0002-7333-5552]{Laura Newburgh}
\affiliation{\YU}
\author[0000-0002-2465-8937]{Anna Ordog}
\affiliation{\UBCO}
\affiliation{\DRAO}
\author[0000-0003-2155-9578 ]{Ue-Li Pen}
\affiliation{\CITA}
\affiliation{\DI}
\affiliation{\ASIAA}
\affiliation{\CIFAR}
\affiliation{\PI}
\author[0000-0002-9516-3245]{Tristan Pinsonneault-Marotte}
\affiliation{\UBC}
\author[0000-0001-6967-7253]{Alex Reda}
\affiliation{\YU}
\author[0000-0003-2631-6217]{Seth R. Siegel}
\affiliation{\PI}
\affiliation{\MU}
\affiliation{\TSI}
\author[0000-0001-7755-902X]{Saurabh Singh}
\affiliation{\RRI}
\author[0000-0002-1491-3738]{Haochen Wang}
\affiliation{\MITK}
\affiliation{\MITP}
\author[0000-0001-7314-9496]{Dallas Wulf}
\affiliation{\MU}
\affiliation{\TSI}
\correspondingauthor{T. Pinsonneault-Marotte, A. Reda}
\email{tristpm@stanford.edu, alex.reda@yale.edu}

\begin{abstract}
We present the first results of the holographic beam mapping program for the Canadian Hydrogen Intensity Mapping Experiment (CHIME). We describe the implementation of the holographic technique as adapted for CHIME, and introduce the processing pipeline which prepares the raw holographic timestreams for analysis of beam features. We use data from six bright sources across the full 400-800\,MHz observing band of CHIME to provide measurements of the co-polar and cross-polar beam response of CHIME in both amplitude and phase for the 1024 dual-polarized feeds instrumented on CHIME. In addition, we present comparisons with independent probes of the CHIME beam which indicate the presence of polarized beam leakage in CHIME. Holographic measurements of the CHIME beam have already been applied in science with CHIME, e.g. in estimating detection significance of far sidelobe FRBs, and in validating the beam models used for CHIME's first detections of \tcm emission (in cross-correlation with measurements of large-scale structure from galaxy surveys and the Lyman-$\alpha$ forest). Measurements presented in this paper, and future holographic results, will provide a unique data set to characterize the CHIME beam and improve the experiment's prospects for a detection of BAO.
\end{abstract}

\section{Introduction}
\label{sec:intro}

The Canadian Hydrogen Intensity Mapping Experiment (CHIME) is a drift scan radio interferometer operating between 400-800\,MHz in 1024 radio frequency bins. It consists of four parabolic cylindrical reflectors, each 20\,m wide and 100\,m long, oriented such that the long axis is aligned North-South. The cylinders are essentially reflective along their length and parabolic (focusing, with $f/D = 0.25$) along their width; the resulting primary beam has an East-West profile formed by diffraction (width $\gtrsim 1^\circ$, from 800\,MHz) and a North-South profile which is essentially the broad ($\sim 120^\circ$ wide across the band) illumination pattern of the CHIME dipoles reflected to the sky. As a result, the telescope's instantaneous field-of-view (FoV) is a narrow strip on meridian extending nearly from horizon-to-horizon; over the course of a day, the rotation of the Earth passes the entire sky through this strip, allowing CHIME to map the full sky visible from its latitude (49.3215 deg north) each day. Each cylinder is populated with 256 dual-polarization feeds along the central 80\,m of the focal line. It is located at the radio-quiet Dominion Radio Astrophysical Observatory (DRAO) site near Penticton, British Columbia, Canada. Its primary science goals are a measurement of large-scale cosmological structure using the redshifted 21\,cm line of neutral hydrogen \citep{Pen:2009, Chang:2010, Masui:2013, Switzer:2013, Anderson:2018, Wolz:2021} and as a radio transient detector for Fast Radio Bursts \citep{Rane-Lorimer:2017, CHIME:FRB_overview} and pulsars \citep{CHIME:Pulsar_overview}.  

The CHIME experiment design is described in more detail in \cite{CHIMEoverview}, and its analog and digital signal chain closely follows the design prototyped in the CHIME pathfinder experiment \citep{Bandura_SPIE:2014}. Amplified signals from all 2048 analog inputs of the CHIME radio feeds \citep{Deng_clover:2017} are sent along 55\,m of coaxial cable, where they are filtered and further amplified prior to being digitized, channelized, and sent by corner-turn network \citep{Bandura_ICE, Mena_RFoF} to the cosmology backend to be correlated and integrated in a GPU-based correlator \citep{GPU_corr}, or sent to the transient backend for transient searching.

CHIME will map the redshifted 21\,cm emission of neutral hydrogen between redshifts $z = 0.8-2.5$ across the northern sky to probe the evolution of the baryon acoustic oscillation (BAO) scale imprinted on the large scale structure of matter (of which the neutral hydrogen is a biased tracer). Cosmological measurements with CHIME are complicated by the presence of bright astrophysical foregrounds (primarily synchrotron emission from our own Galaxy and extragalactic point sources) which dwarf the faint \tcm signal by up to several orders of magnitude. Discriminating between the foregrounds and signal should be possible by leveraging their spectral differences \citep{moraleshewittseparableforeground2004, tcmreionizationforegrounds, foregroundsubtractrequirementstcm, parsonsdelayspecfilterforeground, 
thyagarajanforegroundswidefieldspectra, dayenuforegroundfilter.500.5195E}. However, any spectral structure intrinsic to the instrument (e.g. antenna gains, the beam response) will be imprinted on the foregrounds as well, so separating out the foregrounds requires a high-fidelity characterization of the instrument. Previous work which simulated a measurement of the \tcm power spectrum, corrupted by per-feed beam-width perturbations, suggests that the primary beam per feed as a function of angle and frequency must be known to much better than 1\% \citep{shaver1999, oh2003, liu2011, Shaw_polmodes}. To-date, measurements of \tcm with CHIME have relied on an aggressive foreground filtering scheme, filtering out low-delay (spectrally smooth) modes of the data dominated by foregrounds. With this scheme, CHIME has achieved detections of \tcm in cross-correlation with eBOSS ELG/LRG/QSO samples at redshifts $0.78-1.43$ \citep{CHIMEdetection}, and with the Lyman-$\alpha$ forest at redshifts $1.8-2.5$ \citep{CHIMELymanDetection}. However, as this filtering discards large scale modes along the line-of-sight, these detections were limited to smaller scales insensitive to BAO.

Recovering sensitivity to the large scale modes relevant to the study of BAO will require an improved characterization of the instrument systematics including the beam response. Achieving the aforementioned level of sub-1\% calibration fidelity is challenging for the CHIME instrument for a variety of reasons: the primary and synthesized beams are broad and as a result confusion noise is high; the telescope is stationary and so scanning sources to build up a beam map is not possible; and its far-field is inaccessible to drones or other artificial sources that might be able to transmit a signal for beam mapping purposes. In addition, the CHIME reflectors have proven to be complicated optical systems; \textit{ab-initio} analytical beam models and electromagnetic simulations must take into account both multi-path internal reflections and direct feed-to-feed coupling to achieve a realistic reconstruction of CHIME beam features, and these effects do not admit straightforward parametrizations. 
As a result, a novel program of CHIME beam measurements has evolved, as described in \cite{CHIMEoverview}: (i) solar measurements \citep{dallassolar}, which can measure the common mode primary beam (i.e., the beam profile effectively averaged over the per-feed variations in the beam) with good signal to noise over the sun's semi-annual declination range $-23.5^\circ < \delta < +23.5^\circ$ DEC, and $\alpha\sim\pm$ 20 degrees in Hour Angle (HA); (ii) source catalog measurements, which can be used to measure the North-South beam shape from transit data; and (iii) the holographic measurements described in this paper. Of these measurements, holography is unique in that it can measure the beam shape along East-West transit trajectories to high signal-to-noise, anywhere there is a bright enough source to target.

In this paper we describe our implementation of the holographic technique for CHIME and present results from 529 observations of six sources, representing only a subset of the holographic measurements taken to date (1800 observations of 24 sources).  With this amount of data, these results provide measurements of the CHIME beam with typical radiometric noise at or below the few percent-level relative to the peak main beam amplitude. We identify and present a variety of features in the beams, including the per-feed pointing, beam-widths, feed non-redundancy, overall spectral structure of the sidelobes, etc. 
Section~\ref{sec:holo_description} presents an overview of the holographic technique as applied to CHIME and a steerable telescope (the John A. Galt 26\,m telescope at DRAO), including the formalism describing the measurement, a brief overview of the relevant configuration of the Galt telescope and CHIME correlator, and a procedure to estimate our noise floor due to confusion. Section~\ref{sec:observations} describes the holographic observations, Section~\ref{sec:processing} describes the data processing pipeline, Section~\ref{sec:beams} describes the resulting co-added measurements of the CHIME beams, and Section~\ref{sec:validation} describes consistency checks performed on the results. We conclude in Section \ref{sec:conclusion}.

\section{Description of the CHIME Holographic Technique}
\label{sec:holo_description} 
Holographic measurements of a radio dish have a long history in radio astronomy, primarily as a means of measuring the surface figure of a telescope dish to search for deviations from specifications. For a theoretical overview of the technique, see e.g. \cite{1966holotheory, holographyscottryle}; for some examples of applications of the technique to more traditional radio telescopes, see e.g. \cite{holography30m, holographyeffelsberg100m, holographyyebes40m} and the historical account from \cite{baarsholohistory}.  Using holography to map the far-field beam of a cylindrical telescope was first demonstrated on the CHIME pathfinder array \citep{bergerholo}.

Holographic measurements consist of two radio dishes: a `dish under test' whose beam is being characterized and a reference dish. For the holographic measurements of CHIME, the CHIME inputs are under test, and the co-located equatorially-mounted steerable 26\,m John A. Galt telescope is the reference dish (see \citealt{galt_spie_proc} for details on the Galt telescope, as instrumented for holography with CHIME). The Galt telescope tracks a source as it transits through the CHIME beam, and the signals from CHIME and the Galt telescope are correlated in the CHIME correlator. This correlation measures only what is common between the Galt beam and the larger CHIME primary beam, reducing the confusion noise on a given point source. The resulting interferometric data set provides a measurement of the amplitude and phase of the CHIME far-field beam pattern in both co-polarization and cross-polarization, for each of the 1024 dual-polarized CHIME feeds at all 1024 frequencies. This far-field pattern can be transformed to provide a map of the aperture (as would be done for most holography applications). Because the confusion noise is reduced, this technique provides high signal-to-noise measurements of the CHIME primary beam along the East-West trajectory at the declination of the transiting source. This provides an assessment of sidelobe power which is broadly useful for CHIME science; for example, this sidelobe information was used in \citep{farsidelobefrbs} to obtain signal-to-noise ratios for detections of far-sidelobe FRBs.  After applying an overall calibration per observation (see Section \ref{sec:stacking}), the holographic system is sufficiently stable that multiple observations of a single source can be co-added to improve signal-to-noise. Even with co-adding, only about $\sim$20 sources are bright enough to warrant targeting via the holographic technique, thus it cannot be used to fill the North-South beam. In spite of its sparse declination coverage, holography has excellent synergy with independent probes of the CHIME beam, as the polarization, phase, and especially per-feed variation of the beam are hard to measure without holography. 
\subsection{General formalism}
\label{sec:holo_tech}
Holography is an interferometric measurement of the complex (i.e.\ amplitude and phase) far-field response of a radio telescope, obtained by correlating the signals measured by the antenna under-test and the reference dish while observing a common calibrator. In general, the polarised response of the $i$-th CHIME feed (under-test), in polarization $a$ as a function of sidereal time $\phi$ and frequency $\nu$, to the sky can be written as follows:
\begin{equation}
    F^a_i(\phi, \nu) = \int d^2\mathbf{\hat{n}} J^{ac}_i(\mathbf{\hat{n}}; \nu)\epsilon_c(\mathbf{\hat{n}}; \phi, \nu)e^{2\pi i \mathbf{\hat{n}}\cdot\mathbf{u}_i} 
\end{equation}
where $J^{ac}_i$ is the (direction-dependent) Jones matrix of the CHIME feed, including the beam response, $\epsilon_{c}$ is the vector electromagnetic field density defined on the celestial sphere and incident on the feed,  $\mathbf{\hat{n}}$ denotes a position on the sky in the coordinate frame of the telescope, and $\mathbf{u}_i$ is the position vector (in units of wavelength) of the feed measured from an arbitrary origin point (when we take the cross-correlation with the Galt response, the origin point will be irrelevant). One can write an analogous expression for the response of the (reference) Galt telescope, 
\begin{equation}
    F^{b}_{26}(\nu) = \int d^2\mathbf{\hat{n}} J^{bd}_{26}(\mathbf{\hat{n}} - \mathbf{\hat{n}}_{s}(\phi); \nu)\epsilon_{d}(\mathbf{\hat{n}}; \nu)e^{2\pi i \mathbf{\hat{n}}\cdot\mathbf{u}_{26}} 
\end{equation}
where $J^{bd}_{26}$ is the Jones matrix of the Galt telescope and $\mathbf{\hat{n}}_{s}(\phi)$ denotes the direction of source $s$ at time $\phi$, which the Galt telescope tracks. $\mathbf{u}_{26}$ is the position vector (in units of wavelength) of the Galt telescope from the same (arbitrary) origin point.

 For CHIME, we use a linear polarization basis $a, b \in \{X, Y\}$, with $X$ polarization orthogonal to the CHIME focal axis and $Y$ parallel to it. For the Galt telescope, the polarization basis indexed by $b$ is not fixed relative to CHIME due to the rotation of the dish over the course of an observation, however the feed is aligned such that its polarization axes align with that of CHIME when the telescope is pointed to the meridian. The indices $c, d$ belong to an on-sky polarization basis, and the Jones matrices $J_{i, 26}$ then define a mapping between on-sky polarization and the polarization measured by the instruments. 

In CHIME holography, we form the cross-correlation of the $i$-th CHIME feed and the Galt feed. In general this cross-correlation takes the form:
\begin{align}
\nonumber
    V^{ab}_{i, 26}(\phi, \nu) &\equiv \left<F^{a}_iF^{\dagger b}_{26}\right> \\
    &= \int d^2\mathbf{\hat{n}}\int d^2\mathbf{\hat{n}}'J_i^{ac}\left<\epsilon_c\epsilon^*_d\right>J_{26}^{*db}e^{2\pi i \left(\mathbf{\hat{n}'}\cdot\mathbf
    {u}_{i}- \mathbf{\hat{n}}\cdot\mathbf{u}_{26}\right)}
\end{align}
where the product $\left<\epsilon_c\epsilon^*_d\right>$ is the coherency matrix of the sky signal. 

In practice, the experimental setup allows us to simplify this expression considerably. First, we assume that the electric fields from the astronomical source are spatially incoherent; i.e., the coherency term will introduce a factor $\delta(\mathbf{\hat{n}} - \mathbf{\hat{n}}')$ which eliminates one of the integrals. Second, we note that we perform holographic measurements of \textit{point-source} calibrators transiting over CHIME in tracks of constant declination. To CHIME, the source has a time-dependent position $\mathbf{\hat{n}}_s(\phi)$, while the Galt telescope tracks the source throughout the observation. This is expressed as a $\delta$-function spatial dependence in the sky coherency term, i.e.\ we will write $\left<\epsilon_c\epsilon^*_d\right> \equiv T^s_{cd}\delta(\mathbf{\hat{n}} - \mathbf{\hat{n}}_s)$, where $T^s_{cd}$ are the elements of a matrix describing the polarized emission of the point source.  The $\delta$-function collapses the remaining integral to the direction of the calibrator, leaving us with 
\begin{align}
\label{eq:holo_vis_general}
   V^{ab}_{i, 26}(\phi, \nu) = 
 J^{ac}_i(\mathbf{\hat{n}}_{s}(\phi);  \nu)T^s_{cd}(\nu)J^{*db}_{26}(\nu) \\ \times \nonumber
 e^{2\pi i\mathbf{\hat{n}}_{s}(\phi)\cdot\mathbf{u}_{i, 26}} 
\end{align}
where we have written $\mathbf{u}_{i, 26} = \mathbf{u}_i - \mathbf{u}_{26}$. We have dropped the direction-dependence of the 26\,m's Jones matrix because when we collapse the integrals to the location of the source, only the constant boresight response of the 26\,m remains. 
Combining the phase factor with the sky brightness factor $T^s_{cd}$, we can write the holographic response for all polarization products compactly in matrix form as: 
\begin{equation}
    \mathbf{V}_{i, 26}(\phi, \nu) = \mathbf{J}_i(\mathbf{\hat{n}}_{s}(\phi); \nu)\mathbf{T}_s(\nu)\mathbf{J}^\dagger_{26}(\nu) \ . 
\end{equation}
Given the Jones matrix of the Galt telescope and a suitable, well-known calibrator (i.e.\ a well-constrained $\mathbf{T}_s$), one could invert this expression to solve for the Jones matrix of CHIME feed $i$. However, as we  have ignored instrumental noise, this represents only a schematic illustration of the technique rather than a formal estimator for CHIME's Jones matrix. In practice, there are additional confounding factors beyond instrumental noise: among the most pressing is that the Jones matrix of the Galt telescope is difficult to calibrate independently, as its equatorial mount precludes typical polarimetric calibration through measurements of polarised sources over a range of parallactic angles. In the past (for a different receiver system and observing band), electromagnetic simulations have been used to compute the response of the Galt telescope to polarized emission (see \citealt{tomgalt}). For our purposes, however, a holography-based full-Jones matrix calibration of CHIME, which would include the effects of parallactic angle rotation and polarisation leakage / differential gain in the signal chain, is beyond the scope of this paper. Instead, we directly measure and present the products $\mathbf{V}_{i, 26}(\phi, \nu)$; we discuss our ability to reconstruct independent measurements of the CHIME beam from these products in Section \ref{sec:solarcomp}.

\subsection{John A. Galt Telescope and Receiver}
\label{sec:JGtel}
The John A. Galt telescope \citep{tomgalt} is a 25.6\,m $f/D = 0.3$ (focal length 7.6\,m) equatorial-mount telescope located at the DRAO. It is (254, 22, 19)\,m east, north, and above the center of the CHIME array. A custom 400-800 MHz receiver for the Galt telescope was installed at the focus for holographic measurements with CHIME. The receiver system is described in \cite{galt_spie_proc}, below we highlight some salient features. 

The first element of the receiver chain consists of a dual polarization CHIME cloverleaf antenna \citep{Deng_clover:2017} attached to a sleeve-and-choke structure to circularize the beam. The resulting beamwidths range from $1.2^{\circ}-2^{\circ}$ across the band for both polarizations. One polarization is aligned north-south and the other aligned east-west when the equatorial mount is at Hour Angle = 0 (transit) such that the Galt telescope's polarization frame should align with that of CHIME at transit. Because the equatorial mount rotates with the sky, the polarization frame of the Galt telescope will otherwise rotate with parallactic angle relative to CHIME's fixed polarization. 

The feed is designed with an additional port for calibration signal injection from a local noise source. The signal is split and then attenuated with a 15\,dB Pi-type attenuator that is matched in parallel to 50 $\Omega$ to keep the return loss at the signal injection port $<-30$\,dB. The attenuator is followed by a 2\,mm-wide transmission line which terminates, leaving a 0.7\,mm gap between the transmission line and the microstrip carrying the antenna signals. This gap capacitively couples the injected noise signal to the antenna polarization ports at $\sim$-40\,dB. This low coupling ensures the noise temperature from the attenuator at room temperature is small (0.03\,K) when the noise source is off. 

The calibration source is a 51\,dB ENR NC512/12 Noisecom noise module\footnote{\url{https://noisecom.com/Portals/0/Datasheets/NC500REV3\_datasheet\_WEB.pdf}}, powered by a 12\,V precision power supply and thermally regulated to be 50\,C with a variation of 0.1\,C. A gating signal from the CHIME FPGA is sent along a dedicated coaxial cable to the Galt telescope focus and is used to switch the noise source on and off at a cadence appropriate for gating in the CHIME correlator. The gating scheme switches the noise source on for one \SI{5}{s} integration per minute (i.e., the source is on for one out of every twelve integrations). Measurements of Cygnus A transits by the Galt telescope indicate the signal from the calibration source is $\approx$ 20\,K for both polarizations. This noise source was functional during some of the observations presented in this paper, however the calibration estimation remains a work in progress. As a result, the calibration has not been applied to the data presented here, and all data used comes from the 'gated off' position.

\subsection{Holography Acquisition by the CHIME Correlator System} 
\label{sec:input_correlator}

The CHIME correlator has 2048 inputs, distributed over 128 ICE FPGA boards
\citep{iceboard} which digitize and channelize the input signals. These are
alias-sampled at \SI{800}{\mega \hertz} and Fourier-transformed into 1024
frequency channels streaming at a $\SI{2.56}\mu$s cadence. The digital
streams are transmitted over a high-speed network to an array of 256 GPU nodes where their spatial correlation ($N^2$) is computed and
accumulated in time to a cadence of $\sim$\SI{30}{\milli \second}. Following the GPU kernel, an
estimate of the radiometric noise is calculated by differencing even and odd
frames at the \SI{30}{\milli \second} cadence and recorded alongside the visibilities as a weighting dataset. The data is further accumulated
to 5 or 10 second frames (see below) and transferred to a dedicated receiver node where
additional processing, including calibration and compression, is performed
before saving to disk. 

Signals from the Galt telescope are input to the CHIME correlator and undergo
the same channelization and spatial correlation process alongside every CHIME
input. Following the $N^2$ step, the products between the Galt and CHIME inputs
are split into a separate stream and saved at \SI{5}{s} cadence, faster than the \SI{10}{s} CHIME cadence, to avoid smearing the
faster fringe rate of the longer CHIME-Galt baseline. No additional processing
or calibration is performed on the Galt products before they are saved to disk. 

Although we only report measurements of standard radio sources in this paper, the CHIME correlator is capable of gating on the \SI{30}{\milli \second} frames,
making it possible to perform holographic measurements of pulsars with periods $\gtrsim$\SI{300}{\milli
\second} (a period larger than the observing cadence is required to accommodate the case where the number of observed pulses varies from frame-to-frame). Pulsars are attractive targets for holography because we can remove background emission by differencing frames based on the on/off chopped pulsar signal.

Pulsar timing is computed in
real-time to determine the on- and off-gates using timing model parameters
generated for every observation by the \verb|TEMPO2| package \citep{tempo}. The
gated stream can be switched on or off and the gating parameters altered at any
time without interfering with the rest of the acquisition system.

\subsection{Confusion noise} 
\label{sec:confusion_noise}

One of the limitations of CHIME-only techniques to measure its primary beam is the confusion noise associated with the instrument's large instantaneous field of view. Direct observations of calibrator sources quickly become confusion limited off-meridian. The Sun is bright enough to mitigate this problem, but these measurements are only available in the southern sky due to the limited declination range of the Sun. In holography, the effect of confusion is significantly reduced because the correlation between CHIME and the Galt telescope removes all confusing sources which are outside the Galt telescope's much smaller (compared to CHIME) beam. Here we outline a procedure to estimate the remaining confusion noise limit of the holography data, which sets a fundamental signal-to-noise limit for the holographic technique for non-pulsar sources. 

We proceed from Equation \ref{eq:holo_vis_general}, where we will drop the polarization indices, considering scalar quantities (i.e.\ in only Stokes $I$)  instead of the matrix quantities appearing in the fully polarized treatment (i.e.\ following the discussion at the end of Section \ref{sec:holo_tech}, we will replace Jones matrices $J_i, J_{26}$ with primary beam factors $A_i, A_{26}$). Then, a holographic visibility at a single frequency and sidereal time can be written schematically in the following way: 
\begin{equation}
    V_{i, 26} = A_i(\mathbf{\hat{n}}_s)A_{26}T_se^{2\pi i\mathbf{u}_{i, 26}\cdot\mathbf{\hat{n}}_s}
\end{equation}
where $A_i$ is the primary beam response of CHIME feed $i$, $A_{26}$ is the primary beam of the Galt telescope and the other symbols are as defined in Section \ref{sec:holo_tech}.  

To account for the confusing background of unresolved sources within the Galt beam, we model the brightness temperature as a sum over sources (indexed by $s'$ in the following equations) within the Galt beam (``FoV"). The visibility then reads 
\begin{equation}
    V_{i, 26} = \sum_{{s'} \in \text{FoV}} A_i(\mathbf{\hat{n}}_{s'})A_{26}({\mathbf{\hat{n}}_{s'}})T(\mathbf{\hat{n}}_{s'})e^{2\pi i (\mathbf{u}_i - \mathbf{u}_{26 })\cdot\mathbf{\hat{n}}_{s'}} + \text{src}
\end{equation}
where $+ \text{src}$ represents the contribution of a like term from the actual calibrator source; i.e., the proper holographic visibility we are attempting to measure. The $\mathbf{\hat{n}}_{s'}$ are the sky positions of confusing sources which are near the telescope boresight, i.e., $\mathbf{\hat{n}}_{s'} = \mathbf{\hat{n}}_{s} + \delta\mathbf{\hat{n}}_{s'}$. 

Because the CHIME-Galt baseline fringes rapidly, the geometric phases of the confusing sources are expected to decohere so that the average confusion bias over all sources is negligible. However, the variance of the first term will become a noise term, computed as:
\begin{align}
\nonumber
    \text{Var}[V_{i, 26}]_\text{conf} = |A_i|^2 \sum_{s'}\sum_{s''}A^*_{26}(\mathbf{\hat{n}}_{s'})A_{26}(\mathbf{\hat{n}}_{s''})T(\mathbf{\hat{n}}_{s'})T(\mathbf{\mathbf{\hat{n}}}_{s''}) \\
    \times \left<e^{2\pi i (\mathbf{u}_i - \mathbf{u}_{26})\cdot(\mathbf{\hat{n}}_{s'} - \mathbf{\hat{n}}_{s''})}\right>
\end{align}
where the $A_i({\mathbf{\hat{n}}})$ factor from the CHIME beam has been factored out under the assumption that the CHIME beam varies slowly on scales comparable to the beam width of the Galt telescope.
 As the interferometric phase will decohere for different sources, the ensemble average over the confusing phase term will be non-zero only for $\mathbf{\hat{n}}_{s'} = \mathbf{\hat{n}}_{s''}$, collapsing one of the sums, so that the noise variance reduces to
\begin{equation}
    \text{Var}[V_{i, 26}]_\text{conf} = |A_i|^2 \sum_{{s'} \in \text{FoV}}|A_{26}(\mathbf{\hat{n}}_{s'})T(\mathbf{\hat{n}}_{s'})|^2
    \label{eq:confusion}
\end{equation}
Thus, the confusion noise can be taken as the sum of the squares of the beam-weighted fluxes of the confusing sources within the Galt telescope's field of view.

To estimate this quantity, 1000 realizations of a point source sky were simulated in a HEALpix map ($N_\text{side} = 256$) using the open-source radio sky simulation package \textit{cora} \footnote{\url{https://github.com/radiocosmology/cora/tree/master}}. The simulated point source sky has three ingredients: a Gaussian distributed background of dim sources below 0.1 Jy, a synthetic population drawn using the source count model of \citep{diMatteo} between 0.1 Jy and 10 Jy, and a catalog of real sources drawn from NVSS and VLSS above 10 Jy \citep{nvss, vlss}. These nominal flux thresholds are defined at 151 MHz and the latter 10 Jy threshold for real sources is scaled to the target frequency of the simulation. For each holography source, Equation \ref{eq:confusion} was then evaluated by summing over the squared beam-weighted fluxes of the sources taken within a ring centered on the location of the holography source and of diameter 2 times the full-width at half-max (FWHM) of the Galt beam, using the beam model described in \cite{galt_spie_proc}. We take the square root to obtain a quantity dimensionally compatible with the source flux, and we ignore the prefactor of the CHIME beam as there will be a compensating factor multiplying the source flux.  

The results of this procedure are summarized in Table \ref{tab:confusion} for 600\,MHz. For most holography sources, with the exception of 3C84 and 3C10C which are near other real catalog sources, the results are dominated by the random, synthetic background of sources, which are effectively spatially isotropic. For these holographic sources, over the ensemble of 1000 simulations, the confusion noise from Eq \ref{eq:confusion} has a median value under 1 Jy. As both the source flux and the confusion noise level are modulated by the CHIME beam in an actual measurement, in Table~\ref{tab:confusion} we present the average confusion noise as a percentage of the source flux at 600\,MHz. The reciprocal of this ratio represents the maximum ratio of signal to noise that is achievable with holography; e.g. in our worst case scenario using the very dim source 3C138 ($S_{600} \sim \SI{10}{Jy}$), the confusion noise level is about 10\% of the source flux, such that the peak achievable SNR we may expect is $\sim10$. 
 
\begin{deluxetable}{c c c c}
    \tablecaption{The 24 sources used for holography, their declinations, and their fluxes at 600 MHz, along with estimates for the holographic confusion noise (also at 600 MHz), expressed as a percentage of the source flux. These estimates are defined as in Eq \ref{eq:confusion}.  
    \label{tab:confusion}}
    \tablecolumns{4}
    \tablehead{
        \colhead{Source} & 
        \colhead{DEC [deg]} &
        \colhead{$S_{600}$ [Jy]} &
        \colhead{Conf. Noise [\%]}
    }
    \startdata
    Cygnus A & 40.7 & 3613 & 0.02 \\ 
    Cassiopeia A & 58.8 & 2375 & 0.03 \\
    Taurus A & 22.0 & 1142 & 0.06 \\
    Virgo A & 12.4 & 436 & .2 \\
    Hercules A & 5.0 & 120 & .6 \\
    3C353 & -1.0 & 106 & .7 \\
    Hydra A & -12.1 & 97 & .7 \\
    Perseus B & 29.7 & 87 & .8 \\ 
    3C10C & 64.2 & 71 & 7 \\
    3C273 & 2.1 & 55 & 1 \\
    3C84 & 41.5 & 38 & 10 \\
    3C295 & 52.2 & 37 & 2 \\
    3C58 & 64.8 & 31 & 2 \\
    3C161 & -5.9 & 30 & 2 \\
    3C147 & 49.9 & 29 & 2 \\
    3C111 & 38.0 & 29 & 2 \\
    3C196 & 48.2 & 28 & 2 \\
    3C409 & 23.6 & 27 & 3 \\
    3C358 & -21.5 & 26 & 3 \\
    3C48 & 33.2 & 25 & 3 \\
    3C286 & 30.5 & 18 & 4 \\
    3C454.3 & 16.1 & 12 & 6 \\
    3C78 & 4.1 & 11 & 6 \\
    3C138 & 16.6 & 10 & 7 \\
    \enddata
\end{deluxetable}

\section{Observations}
\label{sec:observations}

In a holographic observation, the Galt telescope tracks a particular source transiting overhead, typically within $\pm30^\circ$ of meridian. During this time, CHIME continues collecting data normally; the signals from the Galt telescope are treated as another input by the CHIME correlator, so that holography observations do not interfere with normal operation of the telescope.  

The Galt telescope is programmed to track sources ahead of time. Pointing during observations is corrected in real-time using a model of the Galt telescope derived in 1998 based on \SI{1.42}{\giga \hertz} observations for the Canadian Galactic Plane Survey \citep{Higgs:Tapping:2000}. A pointing model specifically for the holographic receiver has not yet been developed and is left to future work, however measurements presented in \cite{galt_spie_proc} found pointing offsets of $\sim$3', which we do not expect to affect the measurements presented here.

\begin{figure}[ht]
    \centering
    \includegraphics[width=1.00\linewidth]{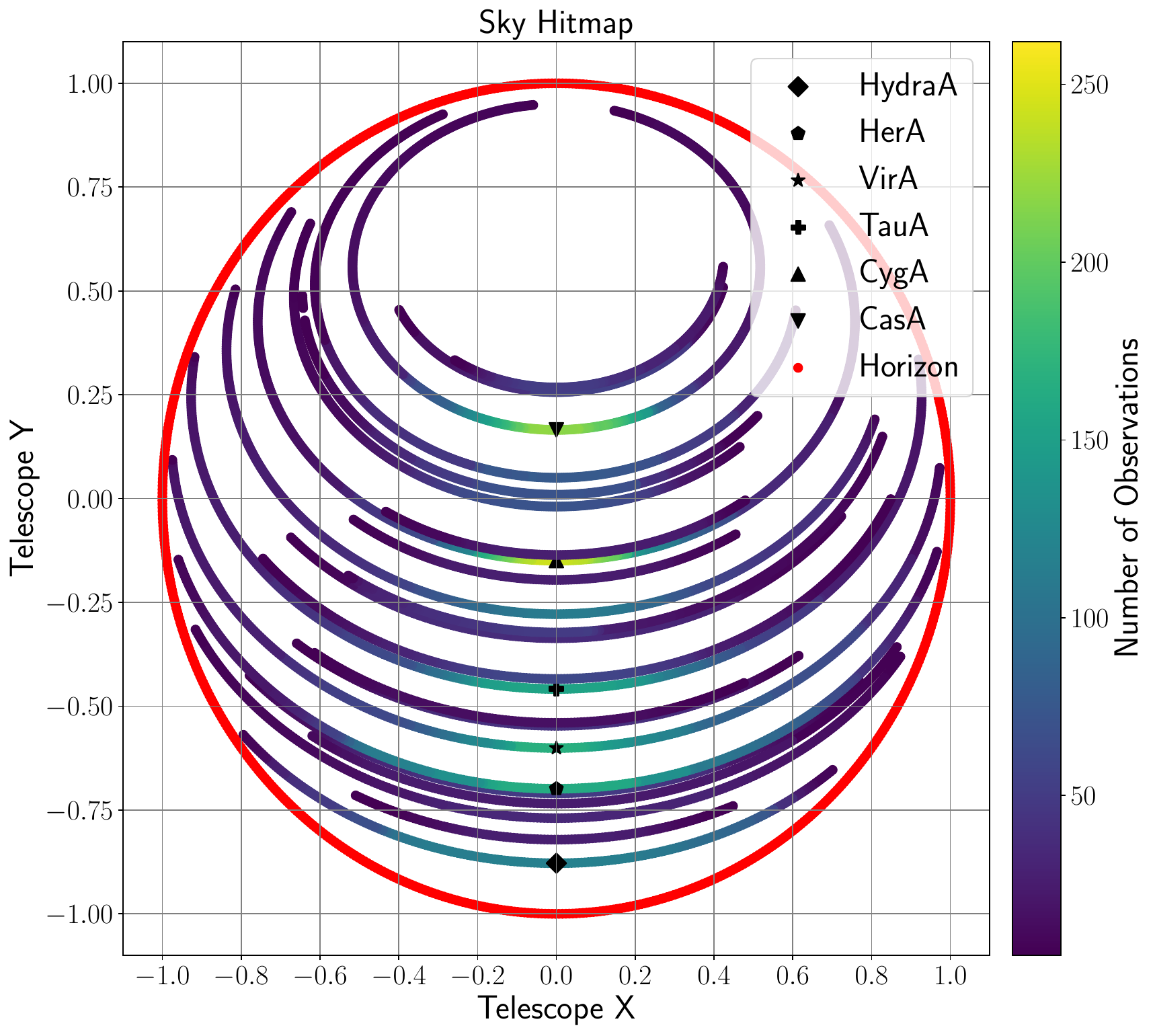}
    \caption{The tracks of all holography observations taken to date, shown in orthographic projection on the sky to illustrate the maximum sky coverage of the dataset. The red circle of radius 1 in these coordinates represents CHIME's horizon; i.e., the inner area of the circle represents the extent of CHIME's observable sky. Various marker shapes denote the centers of tracks that correspond to the brightest sources presented in this work. We note that earlier observations (which tended to be more focused on brighter sources) would track source transits for longer periods of time, corresponding to long tracks on the sky. More recent observations have preferentially (but not exclusively) targeted dimmer sources and in shorter tracks.  }
    \label{fig:alldataonsky}
\end{figure}

In assembling a selection of sources and a schedule for observations, several factors are taken into account. Brighter sources are preferred; for the faintest sources (fluxes on the order of 10\,Jy), in $\approx -20$\,dB sidelobes and with thermal noise at the level of $T_\text{sys} \approx 70$\,K, measuring the beam sidelobes with signal-to-noise of 10 requires in excess of 100 observations. At the same time, observations at a wide variety of source declinations are needed to fill out the 2D beam shape as much as possible. Low declination sources enable a direct comparison with CHIME's solar data-derived beam measurements \citep{dallassolar}, while high declination sources are uniquely observable with holography. Figure \ref{fig:alldataonsky} gives the hitmap of all holography tracks taken to date since October of 2017, illustrating the maximum sky coverage of the current dataset. We use an orthographic projection of the sky onto a frame centered on CHIME with the $y$-axis pointing to celestial North, $x$ pointing East and $z$ pointing to zenith \citep{dallassolar}, and refer to these coordinates as ``Telescope X/Y". This dataset consists of over 1800 individual observations of 24 sources spanning the sky from -21 to +65 degrees in declination. The wide tracks across a large range of declinations offer a unique probe of CHIME's far sidelobes.

\begin{figure*}[ht]
    \centering
    \includegraphics[width=1.0\linewidth]{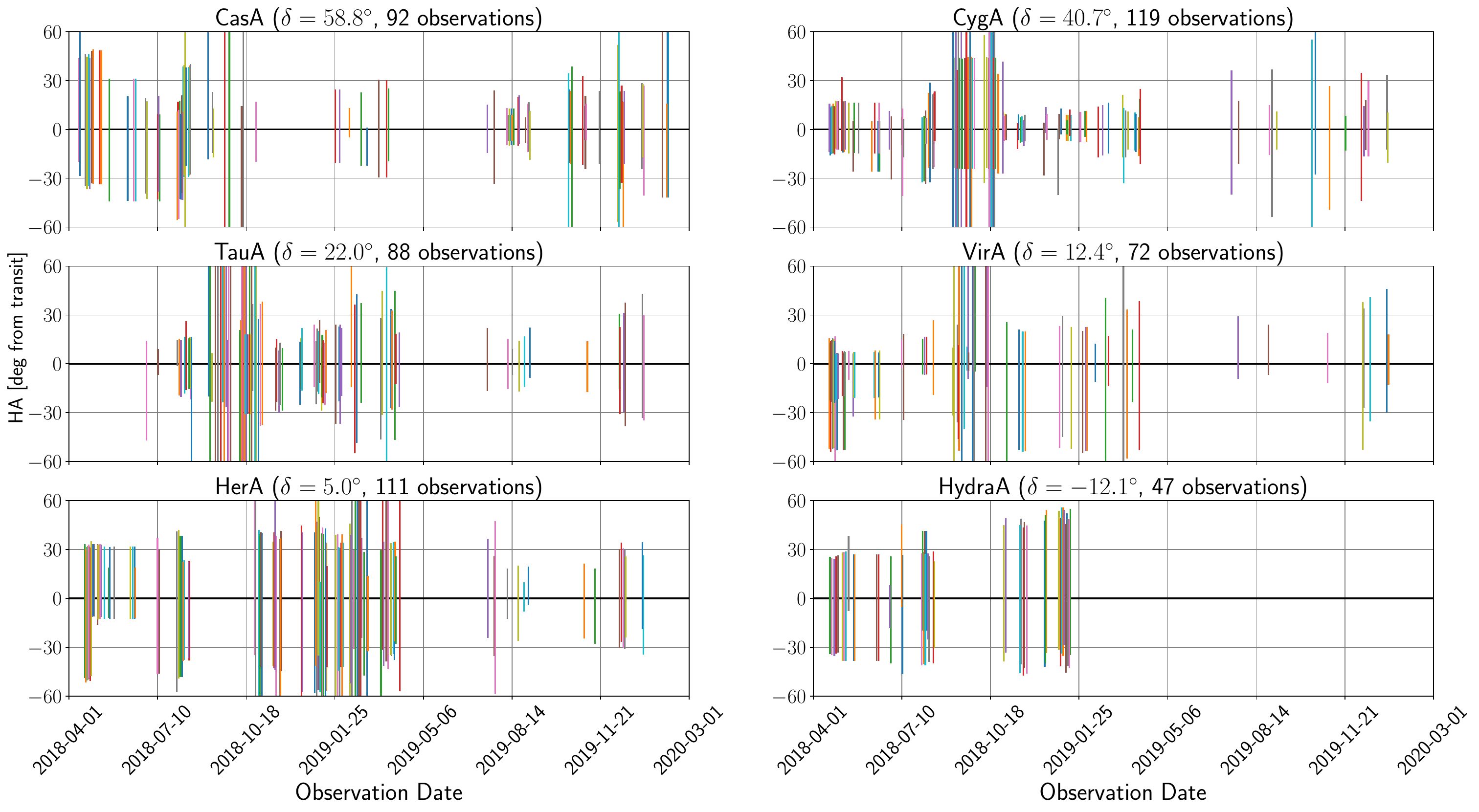}
    \caption{The range of holography observations used in this paper. In each panel the span of the transit in hour angle on a particular day is indicated by the length of the vertical line. Most observations are within $\pm 60^\circ$. The number of observations of each source used in this paper is indicated in the panel titles.} 
    \label{fig:holo_archive}
\end{figure*}

 Figure \ref{fig:holo_archive} shows the time and extent of the holographic measurements presented in this paper, taken between $2018-2020$, of six of the brightest sources available from DRAO: Cygnus A (which may be abbreviated in this work as Cyg A), Cassiopeia A (Cas A), Taurus A (Tau A), Virgo A (Vir A), Hercules A (Her A), and Hydra A. In each panel, the vertical lines denote a single observation of the source as a function of observing date. The vertical extent of the line indicates the span in hour angle of that observation. Typically, the Galt telescope will track a source within $\pm 30^\circ$ of transit over CHIME, but the exact span of the data will vary considerably from transit to transit.

\section{Data Processing}
\label{sec:processing}

In this section we describe the data processing pipeline for holography which interpolates the data onto a common grid in hour angle, corrects for known systematic effects, and averages over repeated observations to improve the signal-to-noise of the beam measurement. 
CHIME data are stored in multiple redundant archives at Compute Canada centres.
Data processing and analysis is mainly conducted on the
Cedar\footnote{\url{https://docs.alliancecan.ca/wiki/Cedar/en}} cluster using a
modular Python-based pipeline and analysis tools developed by members of the
CHIME collaboration\footnote{\url{https://github.com/radiocosmology}, \url{https://github.com/chime-experiment}}.

\subsection{Fringestopping, regridding, decorrelation correction} 
\label{subsec:fringeregriddecorr}

To facilitate the comparison of beam measurements from multiple observations and
sources, the time-ordered holography data are resampled onto a common grid in
hour angle (HA) relative to the transit time of the source on the telescope's
meridian. This regridding is implemented as an inverse Lanczos interpolation, identical to the implementation in the CHIME offline pipeline used in \citealt{CHIMEdetection}. The chosen grid spans $\pm$\ang{30} of hour angle in 720 samples, which is broad
enough to cover most holography observations and sets the resolution of our beam measurement to approximately 0.1$^{\circ}$ in hour angle, taking the data from a \SI{5}{\second} cadence to approximately
\SI{20}{\second}. For
comparison, the minimum full width at half maximum of the CHIME primary (voltage-)beam is 2$^{\circ}$ at 800 MHz, taking at least \SI{480}{\second} for a source to transit, and therefore the
\SI{20}{\second} sampling is sufficient.

Prior to downsampling in HA, a phase correction is applied to remove the fringing
in the data produced by the changing delay between the CHIME and Galt feeds
as the source moves overhead. Without this correction, averaging adjacent time
frames when downsampling would lead to signal loss. The phase correction is 
\begin{equation}
    2 \pi {\mathbf{\hat{n}} \cdot \mathbf{u}_{i, 26}}\text{,}
\end{equation}
the projection of the direction to the source $\mathbf{\hat{n}}$ onto the baseline
between the Galt and the $i$-th CHIME feed in units of wavelength $\lambda$.  For
a source at \ang{0} declination on the meridian, fringestopping corresponds to removing
$>$\SI{4.5}{\radian} of phase over a period of \SI{20}{\second} at 600 MHz.

The amplitude, but not phase, of the raw holographic measurement is affected by an hour-angle dependent multiplicative bias principally generated by cable delays. CHIME contains 55\,m of coaxial cable, while the Galt telescope signal traverses 405\,m of coaxial cable en-route to the correlator. The result is that the signal from the Galt telescope is delayed relative to CHIME's signal by at least 1.49\,$\mu$s, which is a significant fraction of the 2.56\,$\mu$s integration frame. This leads to a significant decorrelation as the signals from CHIME and the Galt telesope no longer overlap within the expected geometric delay. The amplitude of the decorrelation depends on the exact delay between the two signals, which changes over the course of an observation as the source transits, so that the result is not a constant overall suppression of the signal but instead a modulation of the amplitude of the holography measurement as a function of hour angle, distorting the inferred beam shape. 
Fundamentally, the decorrelation is a result of the finite window of time used when channelizing the data in the CHIME correlator's polyphase filter bank (PFB). By simulating the response of the PFB, we can determine an estimate for the amplitude of the decorrelation effect for arbitrary delays and correct the holography amplitude accordingly (see Appendix \ref{sec:decorr_appendix}). Broadly, we find that for the range of hour angles used in the re-gridding step, $\pm30^\circ$, and for the sources considered in this work, this effect modulates the beam amplitude by as much as $\pm\sim30\%$ and the magnitude of the decorrelation increases approximately linearly through the transit (we use the full result, not an approximation, when correcting the data). Referencing to the first feed on each cylinder, there is also a linear dependence on the amplitude with feed, with the sign of the slope dependent on whether the source is north or south of CHIME's zenith (in this paper, the only source north of zenith is Cas A). See Figure \ref{fig:decorr-dependence} in the Appendix.  

As the fringestopping and decorrelation correction steps require knowledge of the geometric delay between a CHIME input and the Galt telescope we must have accurate knowledge of their (relative) positions. Any error in the baseline distances $\mathbf{u}_{i, 26}$ will introduce a phase error in the beam
measurement by biasing our estimate of the geometric delay associated with a source location, but this can be identified and corrected in later stages of
processing if necessary. For instance, an earlier version of the pipeline omitted the vertical component ($w$-term) of the holographic baselines when fringestopping. \cite{galt_spie_proc} details an analysis of that version of the data to fit for the $w$-term, including the contribution of an additional declination-dependent term originating from the geometry of the Galt telescope mount. This model tracks the remaining delay term to within $.1\%$ for 7 sources spanning $>60^\circ$ of declination. We use the best-fit vertical displacement from this analysis for fringestopping in the present version of the pipeline, and include the Galt-specific term as an additional phase correction for the holography prior to averaging over repeated transits (see Section \ref{sec:stacking}). We also determine a best-fit value for the cable delays and use these when applying the decorrelation correction.

In \cite{CHIMEoverview}, we reported evidence of an error in the North-South positions of the CHIME antennas, based on the observation of a phase that linearly scales with East-West position when beamforming to bright point sources on meridian. At the time, we remarked that this could be caused by either a rotation of the telescope from astronomic North, or by a small, linearly staggered North-South offset of the cylinders from one to the next (i.e., an observer looking down at the telescope from zenith would see a parallelogram with the top and bottom sides sloped, rather than a rectangle). In the interim, in all of CHIME's analyses which required knowledge of the baseline distances,  we assumed the rotation scenario; however, a recent professional survey of the telescope confirms the parallelogram scenario. As a result, the baseline distances used here contain an erroneous East-West component which linearly scales with North-South position along the cylinders, which was intended to correct for the rotation scenario. From end-to-end of a cylinder, this makes a difference of about 10\,cm. This error sources a phase gradient in the fringestopped data proportional to $\sin{\text{HA}}$; for Cygnus A at declination of $\approx41^\circ$, at the extent of our grid ($\pm30^\circ$)  this is as much as $23^\circ$ of phase error at 600\,MHz. The delay error sourced by a 10\,cm offset, however, produces no appreciable change in the strength of the decorrelation effect. This will be corrected in future iterations of the pipeline; for now we retain it as it makes no difference to the amplitude and the same phase error is present in the independent CHIME datasets we compare to in Section \ref{sec:validation}. Also note that the previously described phase analysis in \cite{galt_spie_proc} is unaffected by this issue because it used data taken on meridian, where the fringe phase due to the East-West component of the baseline is identically zero. 

\subsection{Data flagging} 
\label{subsec:RFIflaggingtxt}
\subsubsection{Manual data cuts}
Prior to any further processing, we visually inspect all observations used in this paper for overall quality. We completely omit any observations with significant outstanding features from the stacks presented here. Among the features we found which disqualified observations are: 
\begin{itemize}
    \item Significant RFI.
    \item Significant disparities between, or anomalous features in, one or more cylinders.
    \item Large chunks of missing data.
    \item Presence of the Sun at any point in the transit. This is more typical among earlier observations, and is specifically an issue for the southern sources (Tau A, Virgo A, Hercules A, Hydra A). This is a conservative cut as in principle the Sun can be masked when it overlaps the source's track. However, as the overall quality of the data near the Sun is uncertain, we simply omit these data.
\end{itemize}

In total, these by-hand cuts remove 112 of the 529 transits (about 21\% of the data) used for this work. Over half of the discarded transits are of Hercules A and Hydra A, primarily due to solar contamination. 
\subsubsection{RFI flagging}
Although CHIME is located in a radio-quiet zone at DRAO, there is still contamination observed in the data over significant portions of the band due to radio interference from a variety of artificial sources, including satellites, airplanes,  wireless communications, and TV bands \citep{CHIMEoverview}. The Galt telescope sees the same RFI environment, and so holography data is also compromised by this interference. To account for this, prior to fitting and stacking over observations, we apply an RFI mask to the holography data. When available, we use the RFI masks produced by the CHIME offline pipeline, as described in \cite{CHIMEoverview}. 

The CHIME masks are produced for the entire sidereal day; to mask the holography, we select the daily data flags in a window around the transit of the holography source, corresponding to the time span of the observation. Individual time samples in the holography data are then flagged based on a nearest-neighbors interpolation along the time axis of the CHIME masks; i.e. the holography is flagged based on the nearest time sample in the daily mask. We note that some of the holography data presented in this paper was taken at a time before the offline analysis pipeline had been fully developed and so daily RFI masks may not be available for all holography observations. When this is the case, we opt instead to use a static RFI mask which flags known persistent bands of RFI but is not otherwise resolved in time.  

The daily masks are intended to flag anomalous, RFI-contaminated samples in the CHIME data and are not specifically tailored to holography. However, in general we do not expect the RFI in the DRAO environment to decorrelate between CHIME and the Galt telescope. Therefore, to be conservative, we adopt the daily masks to omit data wherever the masks indicate that CHIME was compromised by RFI.

\subsection{Stacking} 
\label{sec:stacking}
To integrate down the noise in our observations, we co-add all available holographic observations of a particular source together to form a lowest-noise estimate of the measured response along the source track, at all frequencies, all feeds, and in all polarization products. We compute the stacks as a weighted average over observations, where the weights are taken as the inverse of the fast-cadence estimate of the noise variance computed by the real-time pipeline as described in Section \ref{sec:input_correlator} (i.e., we perform an inverse-variance weighted average).

\subsubsection{Transit normalization}

Prior to averaging, we must account for variations in the instrument gains between transits. To do this, for every transit, we fit a simple model for the fully complex holographic response in the main lobe, for all frequencies and feeds, in both co-polarization ($YY, XX$) products only; we do not attempt an independent fit of the cross-polarized ($XY, YX)$ response. The amplitude of the response is modeled as a Gaussian parameterized by amplitude $A$, centroid $\mu$ and width $\sigma$. The phase is modeled as a 5th degree polynomial in hour angle $\phi$, $\sum_{i=0}^5c_i\phi^i$. Prior to fitting, we apply the phase corrections discussed in Section \ref{subsec:fringeregriddecorr} and RFI masks as discussed in Section \ref{subsec:RFIflaggingtxt}. When averaging over observations, we then normalize each transit by its measured complex gain, which we take as the best-fit model evaluated on transit (i.e., at $\text{HA} = 0$); in terms of the best-fit parameters, this is given by $Ae^{ic_0}$. This normalization is applied in both co- \textit{and} cross-polarization, so that the amplitude and phase of the cross-polarized response are now referenced to the on-meridian co-polarized response. All best-fit parameters are saved to disk for offline analysis; see Section \ref{sec:beams}.

As this procedure independently normalizes the data for all feeds, frequencies, polarizations, and sources (declinations) to $1 + 0j$ at transit, we are losing information about the modulation of the on-meridian beam response as a function of these dimensions of the data. Although we do not attempt this for the purposes of this work, the dependence of the beam on declination can be re-inserted using independent measurements of (ratios of) the source fluxes, referenced to a specific calibrator source which would provide the overall normalization of the beam. 

\subsubsection{Accounting for noise-source installation}

The procedure described above applies the best-fit gains to both the co- and cross-polarized responses; in the cross-polarized response, this normalization leaves behind a factor of the ratio of the gains of the $X$- and $Y$- polarized inputs of the Galt telescope. We noticed that the cross-polarized holography normalized in this way exhibits a phase jump for all observations following March 8, 2019. This corresponds to the date on which the noise source described in Section \ref{sec:holo_description} was installed on the Galt receiver. This implies that the phase of this ratio, while otherwise stable in time, was impacted by the noise-source installation modifying the receiver signal chain. This phase jump will cause the cross-polarized response to decorrelate in a simple average of observations before/after the noise source installation, so we must first align the phases by applying a correction to the cross-polarized data of all of the observations taken before (or after, as the absolute reference of the phase does not matter) March 8, 2019. To do this, we split the data into two sets, one taken before the noise source was installed, and one after. Using the noise weights, we compute the stack over each of these two sets independently, then fit for the on-transit phase of the cross-polarized data in each stack. We apply the difference of the two phases as a correction to the data taken before the noise source was installed; the result is that all observations are aligned to the current instrumental phase, post-noise source installation. This procedure is carried out for each of our 6 sources, with the exception of Hydra A, for which there are no observations, within the range included in this work, that were taken after March 8, 2019.

\begin{figure*}[ht!]
    \centering
    \includegraphics[width=1.0\linewidth]{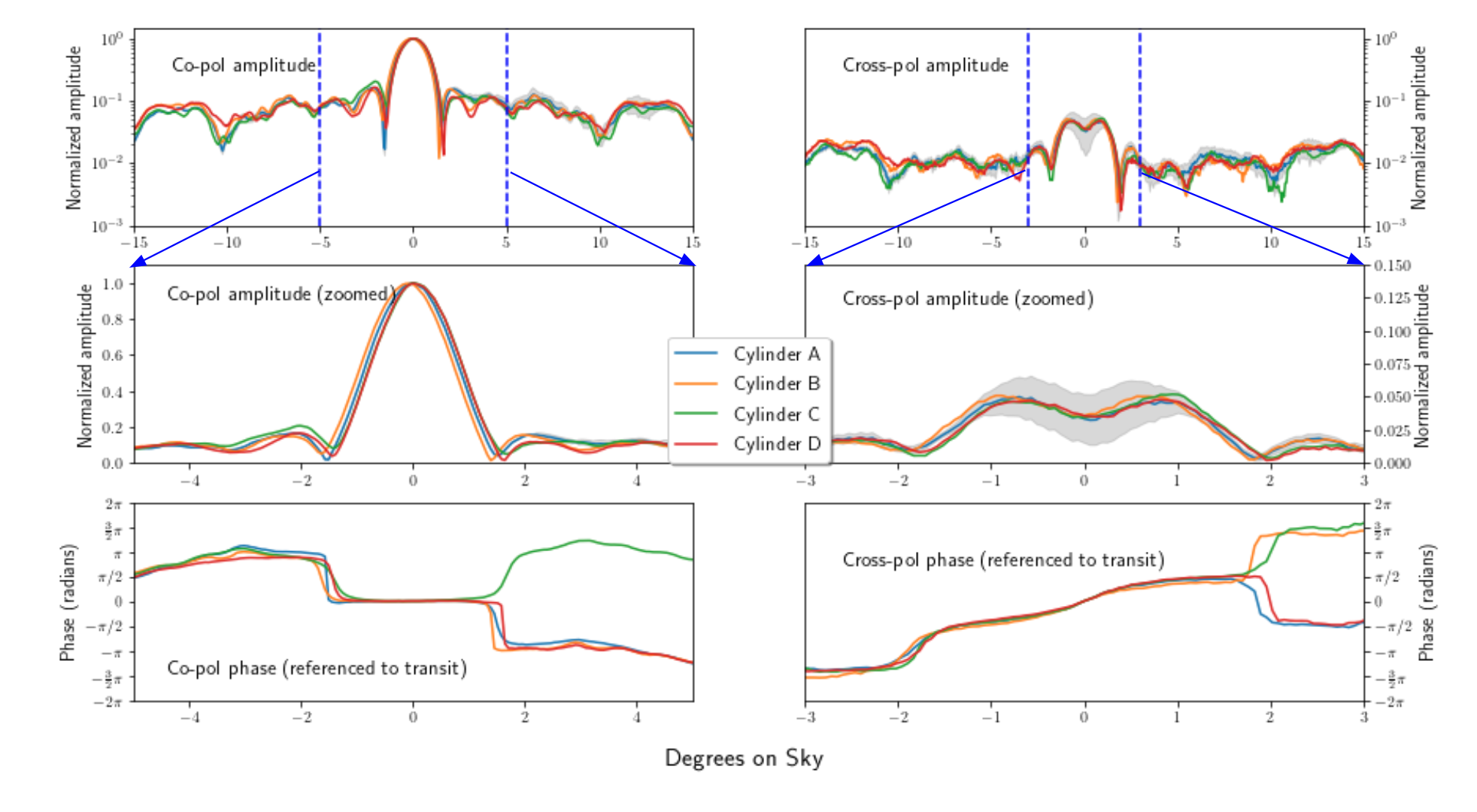}
    \caption{The CHIME $X$-polarized  beam response at \SI{717}{\mega\hertz} from the holographic measurements of a Cyg A transit, taken on 2018 Sep 28. Each panel shows the median response taken over all feeds within a cylinder, normalized such that the co-polar response is $1 + 0j$ at transit.  \textit{Upper left}: The median co-polar amplitude of all feeds per cylinder, normalized by the Gaussian-fit peak height, over the full extent of the observation, converted to degrees on sky, HA$\cdot\cos\left(\delta_{\text{CygA}}\right)$; \textit{Upper right}: The median cross-polar amplitude from the product of a CHIME feed with the opposite polarization on the Galt receiver. The data have been scaled by the same factor as applied to the co-polar response, so the curves give an indication of the level of cross-polarization in the beam; \textit{Middle left}: Same as the upper left panel, but zoomed to a smaller hour angle range and plotted on a linear scale; \textit{Middle right} Same as the upper right panel, but zoomed to a smaller hour angle range and plotted on a linear scale; \textit{Lower left}: The median co-polar phase as a function of scaled hour angle, taken over all feeds in a cylinder (the median was evaluated for the real and imaginary parts separately before evaluating the phase); \textit{Lower right}: Same as the lower left for the cross-polar phase.  The phase difference between cylinders, after accounting for phase wrap, is only large near the first zero crossing of the field.  The gray bands in the amplitude plots indicate the standard deviation over all the Cyg A holography tracks of Cylinder A's median feed response (Cylinder A is representative).}
    \label{fig:holo_overview}
\end{figure*}

\subsubsection{Final stacking}
Finally, the split-stacks are combined into a single, final noise-weighted stack; the noise-weights are propagated to give an estimate of the radiometer noise in the final stack. The stack is saved with 
a record of the number of observations included in each frequency/feed/polarization/hour-angle voxel (this number is not trivially equal to the number of input transits owing both to gaps in the data from masking and the varying lengths of each observation), as well as the empirical variance of the ensemble.

\section{Beam Measurements}
\label{sec:beams}

In this section we present and summarize some of the noteworthy features of the fully processed holography dataset. Each observation presented here is an estimate, as a function of frequency, hour angle, and polarization product, of the per-feed beam response, normalized as described in Section \ref{sec:stacking}; moving forward we denote these quantities as $V_{i, 26}^{ab}$, with $i$ indexing the CHIME feeds, $a, b \in \{X, Y\}$, and where we choose the letter $V$ as a reminder that we are measuring a holographic \textit{visibility} as a proxy for the underlying beam shape. We note that as an interferometer, CHIME's fundamental data products are the correlations between feeds; i.e. the scale of the \textit{power-beam} which modulates the sky as seen by a particular baseline is set approximately (due to feed non-redundancies) by the square of the profiles we show here. When referring to specific polarization products, i.e. $XX, YY, XY, YX$, we use the convention that the first letter indicates the polarization of the CHIME feed, and the second letter indicates that of the Galt telescope.

Some one-dimensional profiles, median-averaged over cylinders, are shown in Figure \ref{fig:holo_overview}, featuring, for Cygnus A, the co- ($XX$) and cross-polarized ($XY$) amplitudes and phases. The profiles are consistent between cylinders, with one of the most notable discrepancies a systematic overall shift of the profile of Cylinder B in the left-middle panel. This suggests that the focal line of Cylinder B is slightly mis-aligned; we return to this point below when discussing the dependence of the beam centroids with feed. The co-polarized response in the main beam is approximately Gaussian, with a first sidelobe around 20\% (4\% in power), and far-sidelobes around $\lesssim3-4$\% ($<$.1\% in power) referenced to the peak. In the main beam, the amplitude of the cross-polarized response is $\lesssim$5\% ($\sim$0.2\%) compared to the co-polarized response (although varies widely with frequency as will be discussed later), and is characterized by a decrement approximately on meridian where the polarization frame of the 26\,m is nominally aligned with that of CHIME. The cross-polarized response in the far-sidelobe region, outside of the blue dashed region in the top row of Figure \ref{fig:holo_overview}, is $\sim$2\% ($<.1\%$ in power), and its shape is qualitatively similar to the co-polarized beam response. The phases of the co- and cross-polarized response vary smoothly within the main lobe and are again consistent between cylinders, except for points corresponding to zero-crossings in the complex beam pattern where otherwise small noise fluctuations may cause the phase to diverge on a different branch (as the ratio of real to imaginary parts is highly sensitive to perturbations near a zero-crossing).

\begin{figure*}[ht]
    \centering
\includegraphics{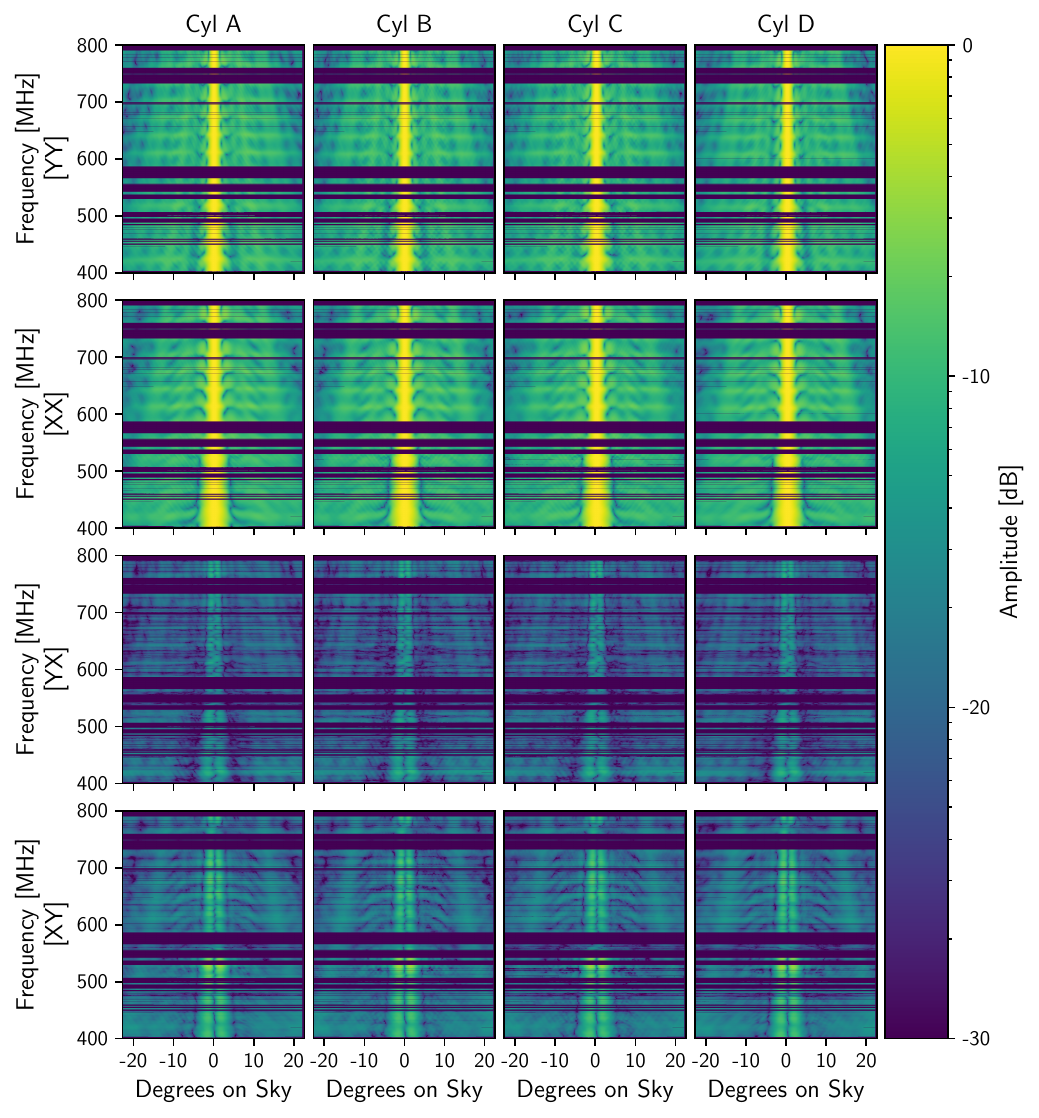}
    \caption{Waterfall plots (in frequency vs sky degrees) of the stacked Cyg A data. For each cylinder, we take a median over all feeds of a given polarization. Zeroed-out bands or points in the data are due to flagged RFI. In this projection of the data the 30 MHz breathing of the beam width (see Figure \ref{fig:fwhms}) creates a ``vertebrae" pattern along the main lobe and first null in the frequency axis. This structure arises from multi-path effects in the cavity between the cylinder vertices and the focal line. The $X$ and $Y$ dipoles of the CHIME cloverleaf feeds do not illuminate the telescope equally; note the change in beam-width and sidelobe patterns between polarizations, as well as an enhanced response in $XY$ relative to $YX$.}
    \label{fig:freqwaterfall}
\end{figure*}

Figure \ref{fig:freqwaterfall} shows two-dimensional slices of the amplitude of the stacked observation of Cyg A, in frequency vs.\ sky degrees, for all four cylinders and all polarization products. Here, we take a median over all feeds of the selected polarization on the respective cylinders. Figures \ref{fig:freqwaterfallCas}, \ref{fig:freqwaterfallTau}, \ref{fig:freqwaterfallVir}, \ref{fig:freqwaterfallHer}, and \ref{fig:freqwaterfallHyd} in the Appendix show the same visualization for the other sources. Superimposed on the expected $\nu^{-1}$ scaling of the main-beam FWHM is an additional widening of the FWHM every 30 MHz in frequency, generating a characteristic``vertebrae" pattern in the main-lobe. This 30 MHz ripple is associated with a standing wave in the cavity between the vertex of the cylinder and the focal line, which are separated by $\sim$ 5 meters as discussed in \cite{CHIMEoverview}. The $X$-polarized response is wider in the East-West direction than the $Y$ response, due to differences in the illumination pattern of the $X$ and $Y$ dipoles of the CHIME cloverleaf feeds; see Figure 9 of \cite{CHIMEoverview}. The side-lobe pattern also differs between the two polarizations; in $YY$ the sidelobes take on a ``checkerboard" pattern while in $XX$ the sidelobes appear relatively smoothed out. The cross-polarized signal is larger in $XY$ than $YX$ by about a factor of 2 in the main-lobe. In all cases, the average beam profile follows the same overall behavior on all of the cylinders.

\begin{figure*}[ht]
    \centering
    \includegraphics[width=1\linewidth]{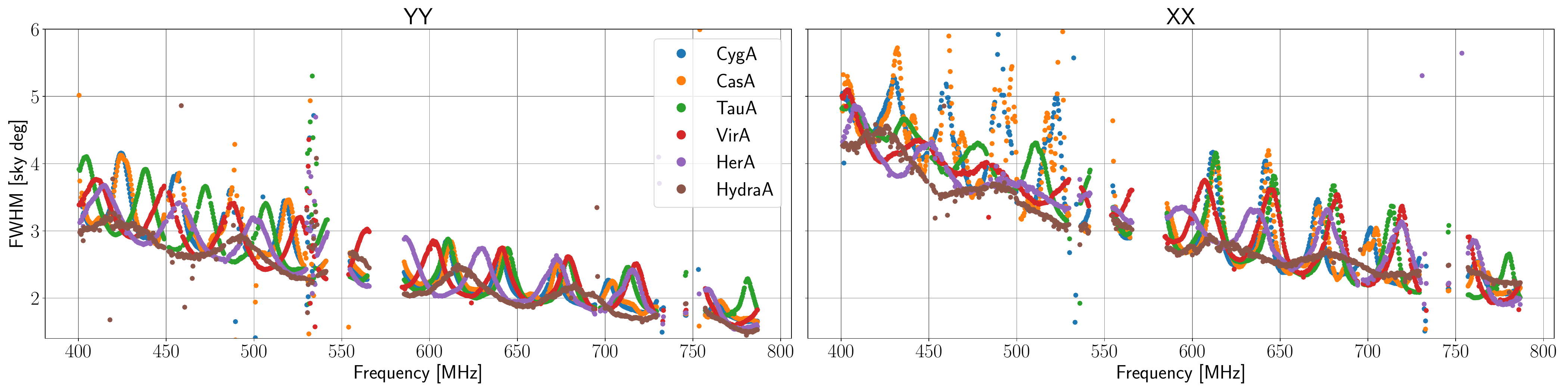}
    \caption{Full-width-at-half-maxima (FWHM) of the holographic beam response as a function of frequency and median-averaged over cylinder A (representative of all cylinders) for all six sources presented in this work. Both polarizations are shown: the X-polarized response is wider in East-West than the Y response, while the 30 MHz ripple (a purely geometric effect) in the beam width is apparent in both polarizations and dominates the modulation of the beam width in frequency. After adjusting for the declination dependence of the on-sky angular distance subtended in 1 degree of hour angle, the magnitude of the beam-widths are consistent between sources, but the locations of the peaks of the 30\,MHz ripple are seen to shift with declination.}
    \label{fig:fwhms}
\end{figure*}

The 30\,MHz modulation and the polarization dependence in the widths is also apparent in Figure \ref{fig:fwhms}, which shows the beam FWHMs -- for all six sources presented in this work -- recovered from the fits prior to stacking as a function of frequency. The beam-widths are consistent between all declinations measured here, although the 30\,MHz ripple patterns are offset. 

\begin{figure*}[ht]
    \centering
    \includegraphics{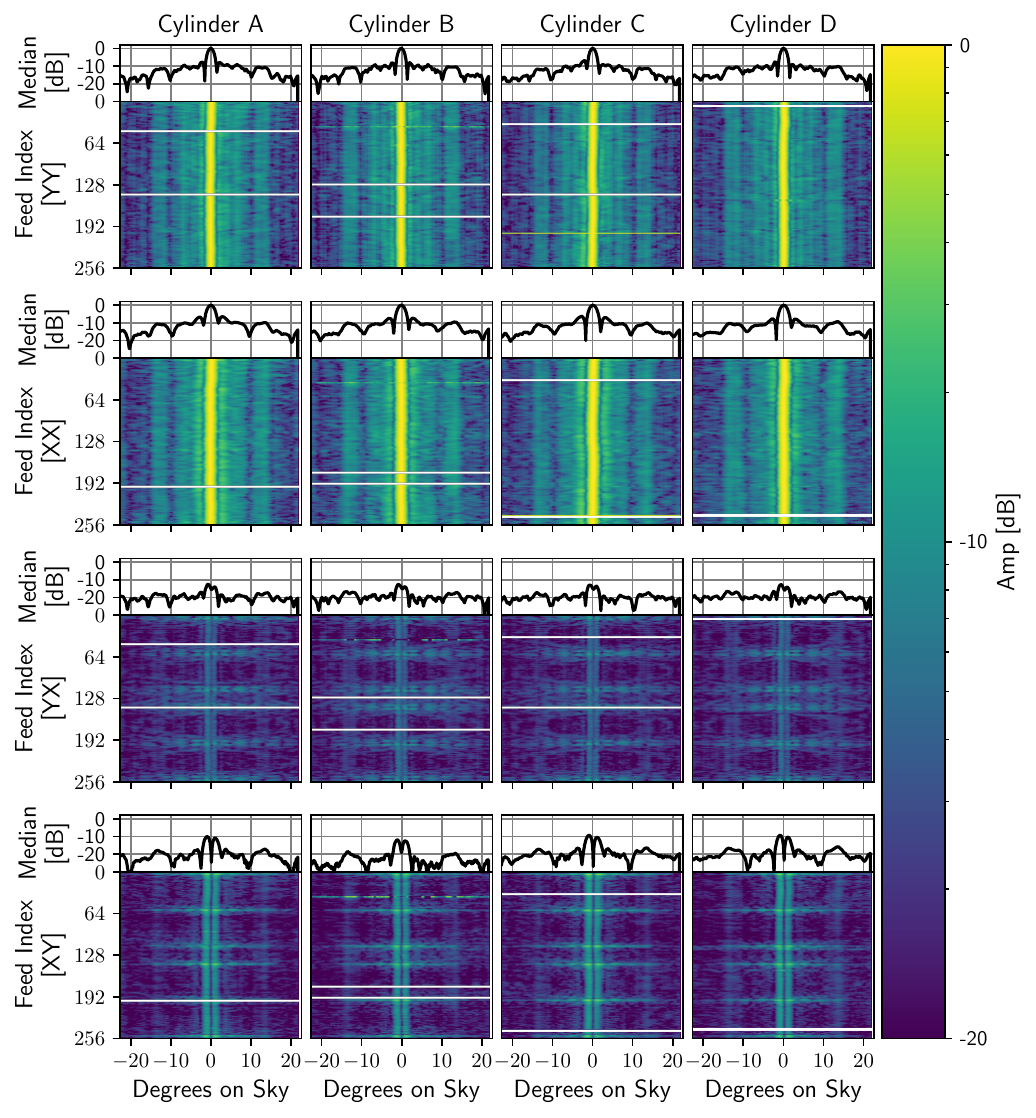}
    \caption{Waterfall plots (in feed index vs. sky degrees) of the stacked Cyg A data. For each cylinder data is taken from 717 MHz. The top two rows show co-pol data while the bottom two rows show cross-pol. The one dimensional slices above each image panel show the median along the respective cylinders. At a fixed frequency, the width of the main beam is stable along feeds, although the beams in X polarization are wider than those in Y. The beam centroids vary along the cylinders due to small misalignments of the feeds from the cylinder axis. There are also asymmetries in the first sidelobes. The cross-pol response exhibits a strong enhancement off-axis at the locations of the focal-line support legs.}
    \label{fig:feedwaterfall}
\end{figure*}

\begin{figure}[ht]
    \centering
    \includegraphics[width=1.0\linewidth]{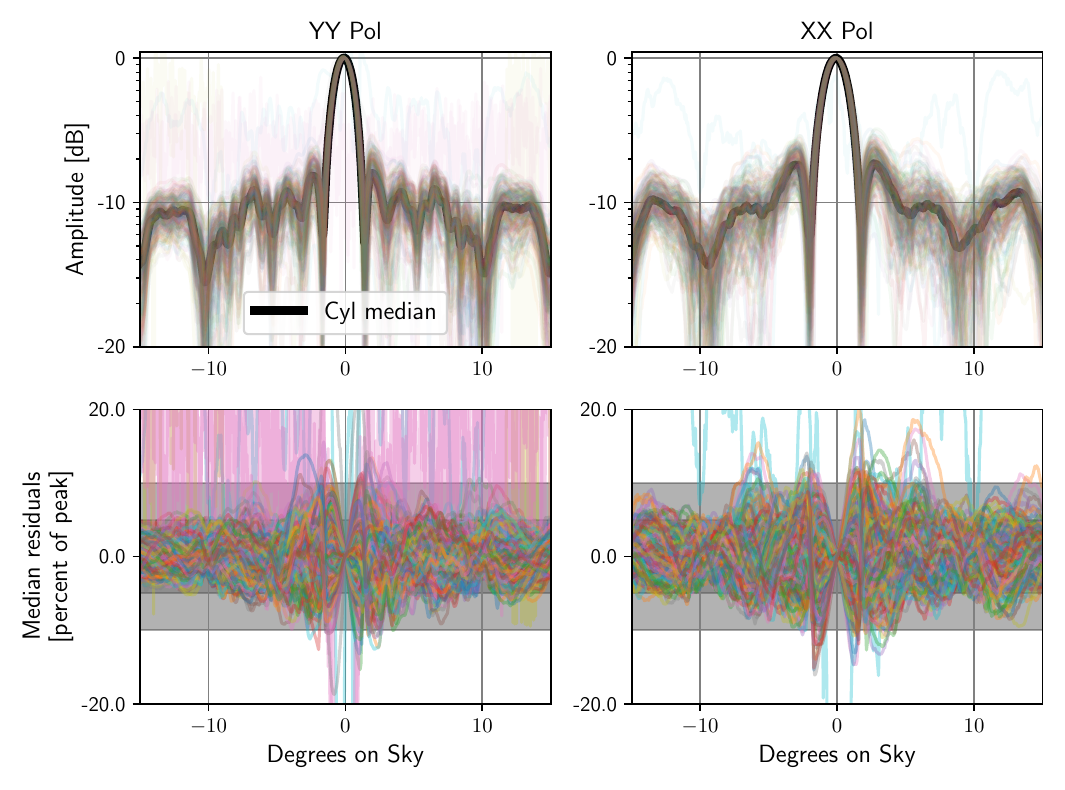}
    \caption{Top: All feeds on Cylinder B, at 717 MHz, overplotted in colors with the median along Cylinder B in black. Bottom: Residuals between the median profile of Cylinder B and all feeds on Cylinder B. The shaded bands around 0 represent 5 and 10\% deviations from the median, relative to the peak. CHIME is a highly redundant array geometrically; here, we observe variations in the per-feed beam response off-axis which are comparable to the overall sidelobe level. The larger residuals near transit are primarily due to centroid wander along feeds; see Figure \ref{fig:centroids}. }
    \label{fig:nonredundancy}
\end{figure}
\begin{figure*}[ht]
    \centering
    \includegraphics{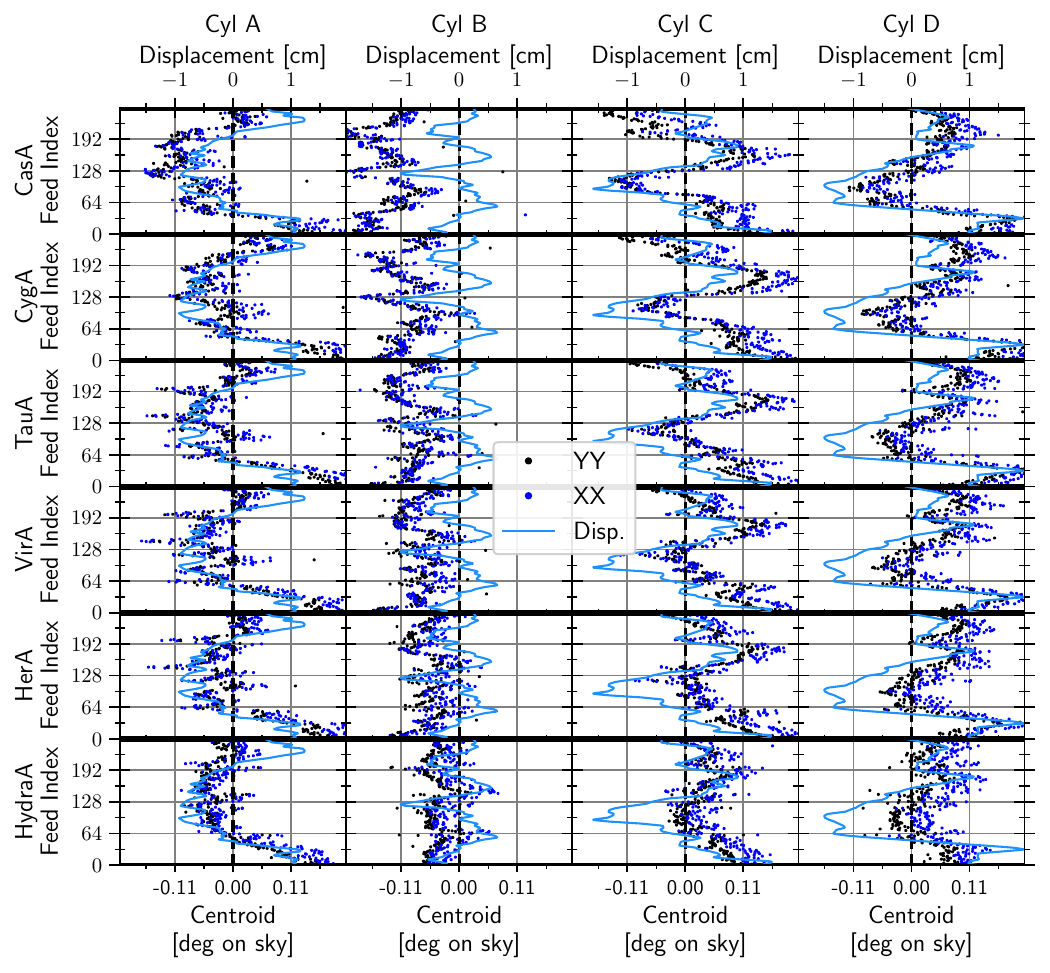}
    \caption{The beam centroids for all feeds along the four cylinders (columns, centroids plotted against the lower $x$-axis), and for all six sources (rows, in descending order from the highest declination source at the top, Cas A, to the lowest, Hydra A, at the bottom). Both polarizations are overplotted, and we take a median average over the band.  The patterns of variation within a cylinder are consistent between polarizations. The centroid wander is mostly within $\pm 0.15^\circ$, corresponding to offsets of the feeds from the cylinder axis $\sim$ 1\,cm. The solid blue lines indicate measurements of the actual physical displacements (plotted against the upper $x$-axis) of the CHIME feeds obtained through photogrammetry of the focal lines; these measurements can only determine the displacements up to an overall misalignment of the focal line and so are always centered about 0. The centroid variation is, as expected, correlated with the pattern of feed misalignments, up to the missing overall offset per cylinder. This overall offset noticeably scales with declination for Cylinder B.} 
    \label{fig:centroids}
\end{figure*}

\begin{figure}[ht]
    \centering
    \includegraphics[width=1.0\linewidth]{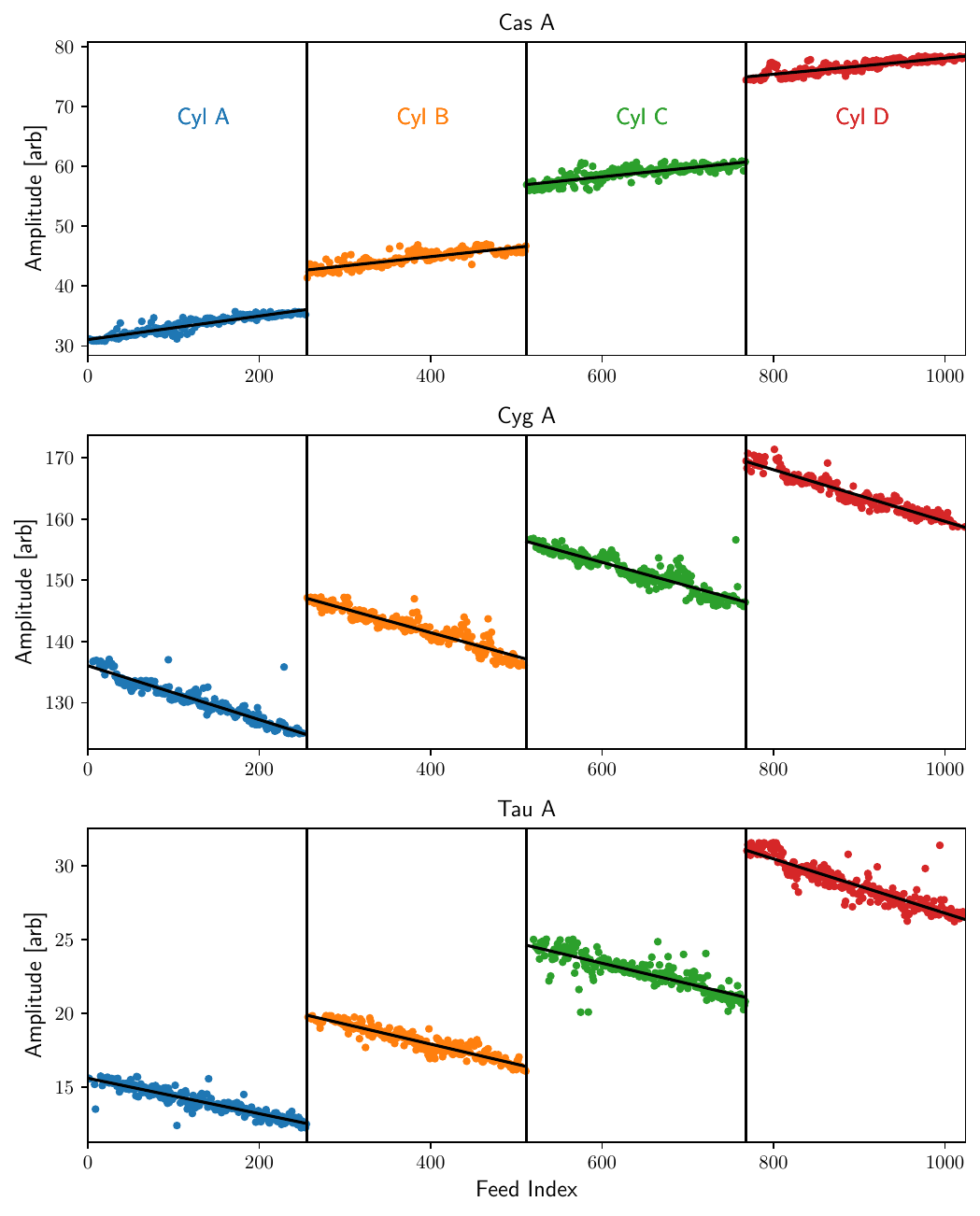}
    \caption{The best fit peak amplitudes in $YY$ polarization of several different transits overplotted for each of Cas A, Cyg A, and Tau A. The four cylinders are plotted in different colors to highlight the discrete jumps in amplitude between cylinders. One expects the response of an interferometer to be independent of baseline for a true point source, however the discrete jumps in amplitude between cylinders indicate that these sources are resolved to an extent by the holographic ($\sim 300\,\text{m}$) baselines. There is an additional overall linear trend in the amplitudes as a function of feed along the cylinders; a best-fit black line guides the eye for each cylinder.}
    \label{fig:amps}
\end{figure}

Figure \ref{fig:feedwaterfall} shows two-dimensional slices of the amplitude of the stacked observation of Cyg A, in cylinder feed-number vs sky degrees, again for all four cylinders and all polarization products. Figures \ref{fig:feedwaterfallCas}, \ref{fig:feedwaterfallTau}, \ref{fig:feedwaterfallVir}, \ref{fig:feedwaterfallHer}, and \ref{fig:feedwaterfallHyd} in the Appendix show the equivalent plots for the other sources. Here, the East-West beam profile in co-polarization is consistent between feeds on a cylinder. Beyond the main beam, the sidelobes plateau at a level around -10\,dB, with the next notable feature an abrupt drop, usually of about 5\,dB across the cylinders and polarizations, at $\approx\pm10^\circ$. The cross-polarization response is shown in the bottom two rows; as before, there is a central null on-transit when the polarization frame of the Galt feed is aligned with that of CHIME, so any signal is the result of genuine polarization either on-sky or inherent to the instrument. We again observe a secondary null at $\approx\pm10^\circ$. We also find that feeds near the legs exhibit an enhanced response off-axis which appears as near-horizontal bands coinciding with the feed legs. The inset-panels at the top of each row show the median profile over the cylinder.

In Figure \ref{fig:nonredundancy} we show an example of the variation in the beam across all feeds on Cylinder B in $YY$ polarization for Cyg A. Within the main beam, deviations are $\sim$10\% of the peak, driven by the feed-dependent centroid wander. In the sidelobes, essentially all (non-outlier) feeds lie within 5\% of the median peak; this corresponds to order unity variation in the sidelobe level from feed to feed.

This centroid wander is one of the dominant modes of variation along the feed axis, as was described in detail in \cite{CHIMEoverview}. In that paper, using holographic data from only CygA, we noted that the centroid wander was suggestive of $\lesssim$1.7\,cm physical offsets between the E-W feed position and the symmetry plane of the cylinders. Here we present the centroids from all six sources along each cylinder, for both co-polarization products, taken as a median across frequencies (although the amplitude of the centroid offset depends on frequency and is also subject to a 30\,MHz ripple, the pattern across the feeds is correlated across all frequencies). The results are shown in Figure \ref{fig:centroids}, where we plot the centroids of both co-polarizations along with measurements of the physical displacements of the feeds taken through photogrammetry of the focal lines. We find that the patterns in the centroids as a function of feed are correlated with the physical displacements (up to an overall misalignment which our photogrammetry is insensitive to), and are consistent between all of the sources, with the exception of those on Cylinder B. It is clear that the overall bias on Cylinder B is declination dependent; the average centroid offset on Cylinder B worsens with increasing declination, changing from an average of $-0.05$ degrees at $\delta=-12^\circ$ to $-0.25$ degrees at $\delta=59^\circ$, which is about 16\% of the (voltage) primary beam width in $YY$ at the top of the band ($\sim1.6^\circ$, see Figure \ref{fig:fwhms}); e.g. beamforming the North-South baselines on Cylinder B to meridian at high elevations would suffer about $\lesssim5\%$ (10\% in power) signal loss due to primary beam attenuation. In \cite{CHIMEoverview}, we measured the overall centroid offset of Cylinder B with holography of only Cygnus A; from that offset we inferred an overall misalignment of the focal line of Cylinder B. While the measurements of Cygnus A here are consistent with that work, the declination dependence that becomes apparent with this expanded dataset is inconsistent with the interpretation of the overall offset as originating purely from a focal line misalignment. This is not a systematic inherent to the holography measurement; because we do not see the same trend in the other cylinders, and an overall linear trend with elevation in the centroids of Cylinder B is also observed in the solar measurement of the beam. The exact cause of this feature remains an area of active investigation.

We also find a dependence of the holography amplitude with feed both along and between cylinders. In Figure \ref{fig:amps} we show the best-fit amplitudes for Cas A, Cyg A, and Tau A (these amplitudes, we reiterate, are removed from the data prior to stacking) for a few transits of each source. We note two features: a strongly linear dependence of the amplitude with feed within each cylinder, and an overall jump in the average amplitude between cylinders. This latter effect can be understood as the result of the holographic baselines resolving out a portion of the extended structure of the source; as we move from longer to shorter baselines (with cylinder A being the farthest cylinder from the Galt telescope and cylinder D the closest) the characteristic fringe pattern on the sky takes on longer wavelengths and thus resolves less of the source, increasing the apparent power. The overall slope within each cylinder, noting that the sign of the slope changes across zenith, is not yet understood. The 26\,m is aligned $\sim20$\,m north of the center of the CHIME array while the portion of the focal lines outfitted with feeds extends $\pm40$\,m from the center; i.e., the baseline lengths between CHIME and the 26\,m do not increase monotonically along a given cylinder. As a result, we would not expect signal loss from resolving the sources to behave linearly along the cylinders. The linear behavior, along with the sign-flip over zenith, is qualitatively similar to the behavior of the decorrelation correction (see Figure \ref{fig:decorr-dependence}), however, the decorrelation correction has the opposite overall sign. This remains an area of active investigation, but we do not expect this to impact the measurements of the beam shape we show here due to our normalization scheme.

Finally, we also find variation in the first-sidelobe levels. We note that for many feeds these sidelobes appear to be asymmetric. In Figure \ref{fig:feedwaterfall} this can be seen both at the per-feed level and in the medians over the cylinders. There are multiple potential sources of asymmetry in the holography data, including an error in the applied decorrelation correction and comatic aberration. Coma (see for instance \citealt{baarsbook}) arises from a lateral misalignment of the feed of a radio telescope from the focal point of the reflector and can generate asymmetric sidelobes. As first shown in \cite{CHIMEoverview} and also described above, we infer deviations of the CHIME feeds from the symmetry axis of the cylinders from measurements of the beam centroids. In principle, this displacement can be used to model a phase error in the aperture illumination and predict the expected sidelobe asymmetry; we leave this to future work. 

\subsection{Aperture transforms}
\label{sec:aperture}
In the usual application of the holographic technique, one uses the phase information of the interferometric measurement to uniquely invert the Fourier transform mapping between the beam farfield and the electric fields in the aperture of the telescope.  

Typically one imagines the aperture plane as the finite two-dimensional surface through which all radiation incident on the telescope must pass; the electric fields across this surface are then Fourier transformed giving the two-dimensional (in terms of sky coordinates) beam response in the farfield regime. For example, for a typical parabolic reflector, the aperture plane is a disk of diameter equal to that of the reflector; if this plane is uniformly illuminated then the Fourier transform of the disk yields the familiar Airy diffraction pattern. 

The case of CHIME is more complicated to treat exactly as there is no finite two-dimensional surface to take for the aperture of the reflector, due to the open ends of the cylinders. Instead, for the purposes of the strictly qualitative overview of aperture features we give here, we take the following approach. We treat the CHIME cylinders as diffractive systems only along the parabolic axis; along the focal axes of the cylinders they merely reflect rather than focus incident light. We then treat the aperture-to-farfield mapping as a one-dimensional Fourier transform; i.e., a transform along the East-West (diffraction) axis. If we consider emission from a source at a fixed incidence angle $\theta$ measured with respect to the local zenith, then the result of our aperture transform can be interpreted as a measurement of the aperture illumination as a function of the aperture coordinate $x$ (centered at 0 on the focal line and extending out in either direction with the cylinder edges at $x = \pm10\,\text{m}$), at a fixed position $y = 2f\tan\theta$ along the cylinder axis.

When analyzing holography data in this context, this is only an approximate mapping; as a source moves overhead, it follows a curved track on the sky so that the incidence angle of its emission on the cylinders is not fixed. Thus for holography the aperture measurement will be affected by variations in the beam over the North-South extent of the track in telescope coordinates, and will also mix power at neighboring positions along the focal line. 

\begin{figure*}
    \centering
    \includegraphics[width=1.0\linewidth]{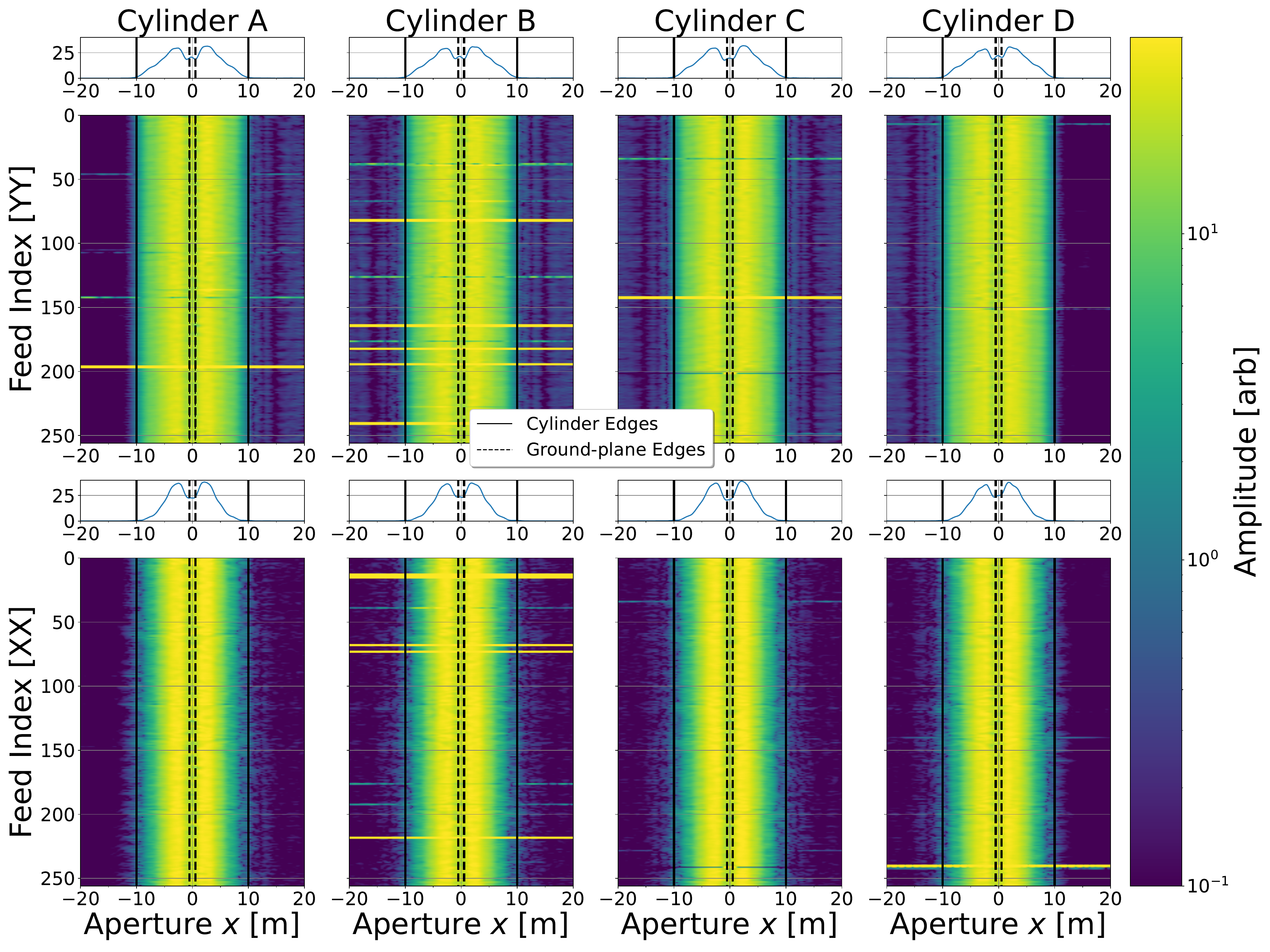}
    \caption{The (amplitude of the) aperture illumination inferred from holography of Cyg A, as a function of $x$ (in meters relative to the focal line at $x=0$\,m) and feed, for both co-polarizations and at 717\,MHz. The black dashed lines indicate the extent of the focal-line ground plane and the solid black lines indicate the physical edges of the cylinders. The pattern is consistent between different feeds; the medians over cylinders are shown in the panels above each image. In $YY$ polarization especially there is an apparent coupling between neighboring cylinders given the absence of power beyond $-10 (+10)$\,m for Cylinder A (D), where there is no other neighboring cylinder. }
    \label{fig:aperture_transform}
\end{figure*}

Figure \ref{fig:aperture_transform} shows the aperture transform of Cyg A as a function of feed along the cylinder and the aperture $x$ coordinate. We find there is little variation in the $x$-profile of the aperture illumination as a function of feed. As a function of $x$, the dominant features are \textit{(i)} a decrement in the central region corresponding to the ground-plane; and \textit{(ii)} evidence for non-zero response beyond the edges of the cylinder, which would indicate cross-coupling between feeds of adjacent cylinders. This is most apparent in the $Y$-pol case; the inner cylinders, B and C, see excess power beyond the physical aperture on either side due to being flanked on both sides by other cylinders, while cylinders A and D have a negligible response on the side opposite of the inner cylinders. Finally, we note that as this is essentially a Fourier transform of the data in hour angle, the erroneous phase gradient in hour angle introduced to correct for the rotated-telescope scenario discussed in Section \ref{subsec:fringeregriddecorr} implies that from end-to-end of a cylinder, the center of the shape of the aperture transform is shifted from its true position in $x$-space by $\lesssim10$\,cm.

\section{Validation and Systematics}
\label{sec:validation}

\subsection{Noise-performance and jack-knives} 
\label{subsec:noiseandjack}
\begin{figure}[ht]
    \centering
    \includegraphics[width=1.0\linewidth]{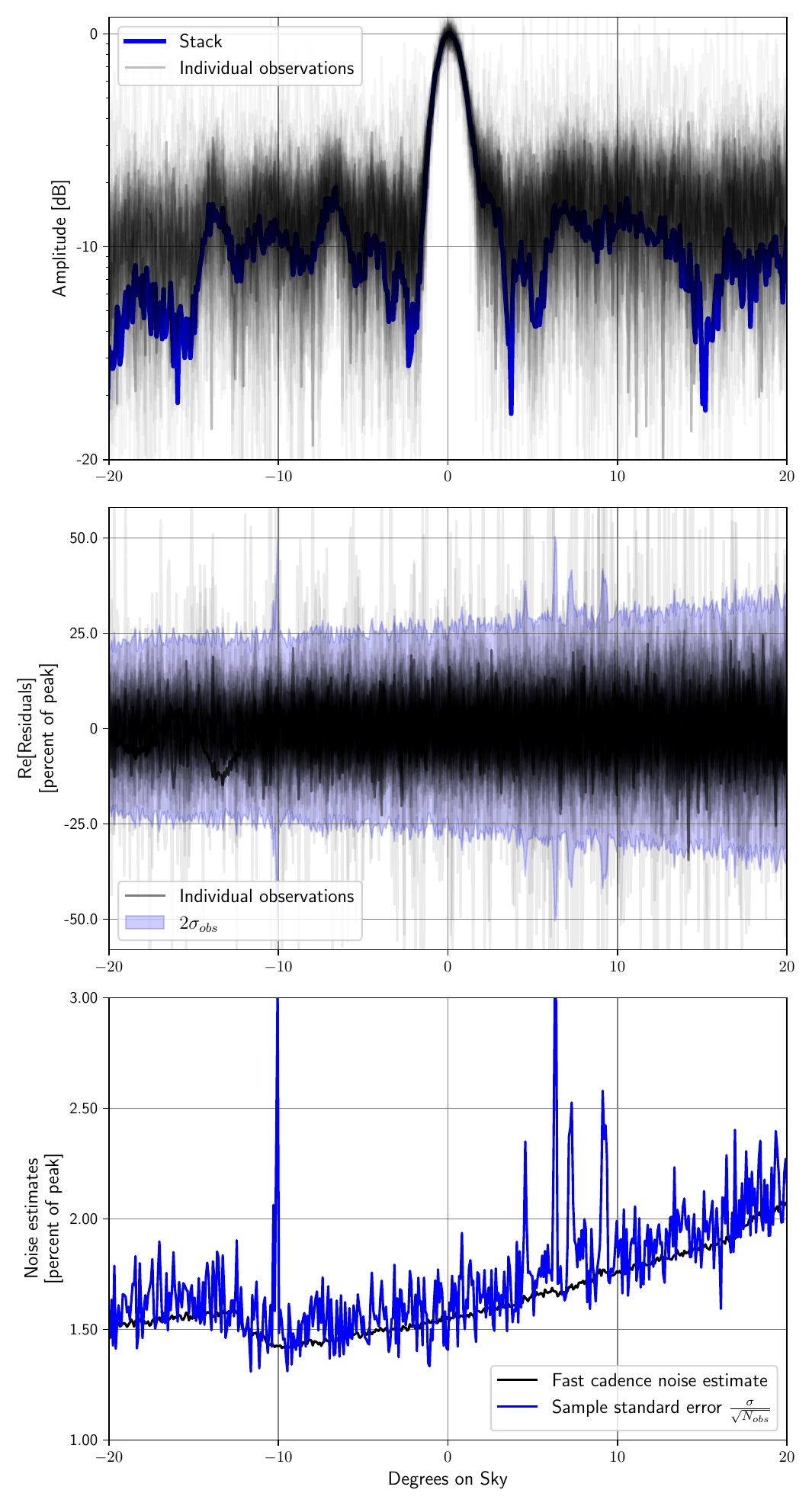}
    \caption{
    \textit{Top panel}: Amplitude of the full stack of Hercules A transits for a single frequency and feed. All input transits are overplotted in gray. The axes are in logarithmic scale. The individual transits are dominated by the noise bias at the level of -10\,dB, while beam features outside the main lobe and at lower amplitudes start to be discerned in the stack. \textit{Middle panel}: The real part of the residuals between the stack and the individual transits, where the blue shaded region indicates the span of two standard errors about the mean $\sigma_\text{obs}$, and centered about 0. \textit{Bottom panel}: The standard error on the stack (in blue) compared to the fast cadence estimate of the radiometric noise (in black), indicating that the noise observed in the stack for this frequency and feed follows the behavior of the expected thermal noise. }
    \label{fig:hercules_stack_noise}
\end{figure}
\begin{figure*}[ht]
    \centering
    \includegraphics{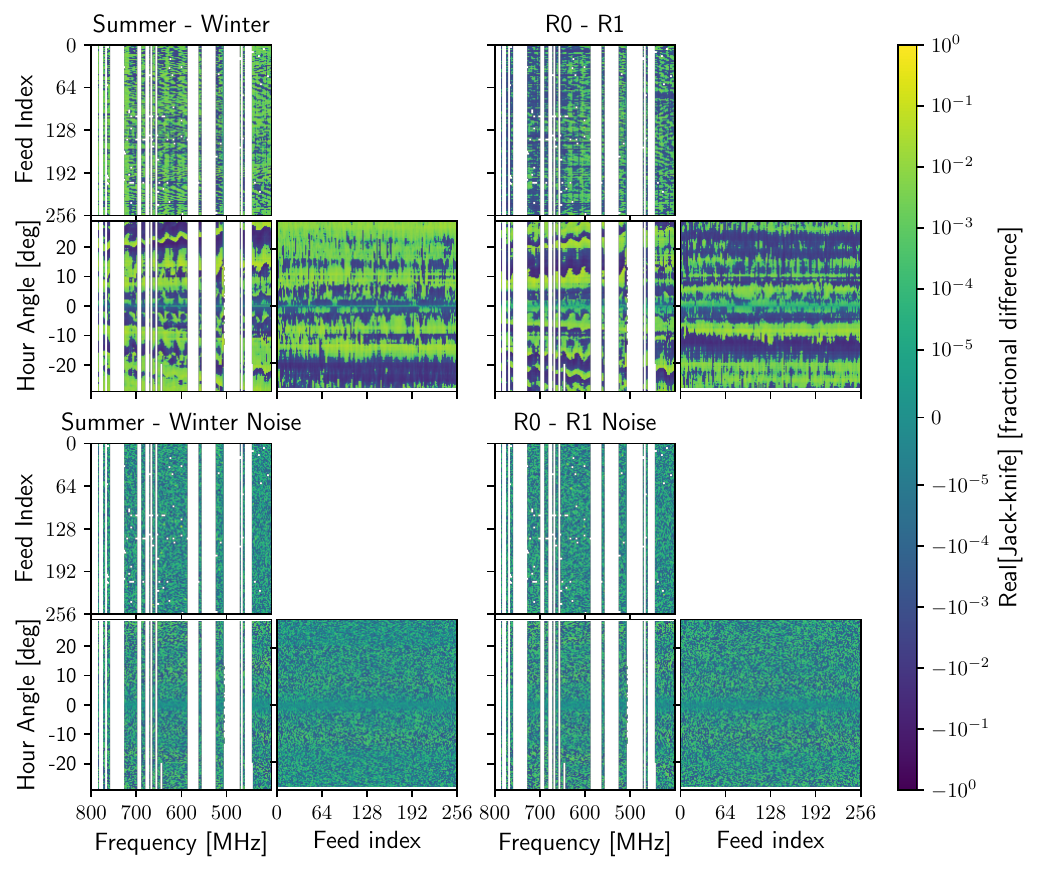}
    \caption{Top row of corner-plots: jack-knives for two different partitions of the Cyg A dataset. The three image panels for each case show the real part of the fractional residuals (i.e., normalized by the value of the beam as measured by the final stack in that pixel) in the three different spaces of the data: feed (for cylinder A) vs frequency (taking a median over hour angle), hour angle vs frequency (taking a median over feeds on cylinder A), and hour angle vs feed (taking a median over frequency).  Bottom row of corner-plots: Noise realizations of the jack-knives, i.e. what we would have expected to see from the differencing procedure had the stacks of each partition been identical with the exception of their noise realizations. In both cases, only the real part is plotted. The residuals indicate systematic differences between partitions of the data  at the sub-percent level between disjoint subsets of the data, however, from the comparison of the two jack-knives it is not evident that these systematics are correlated with temperature.} 
    \label{fig:jack}
\end{figure*}

An example of the final output of the processing pipeline described in Section \ref{sec:processing} is shown in Figure \ref{fig:hercules_stack_noise}, where the stack of Hercules A, for a single feed and frequency in $YY$ polarization, is shown in blue in the top panel with all individual input transits of Hercules A overplotted in gray. We see the intended effect of the averaging procedure reducing the noise in the beam measurement; outside of the main lobe, the individual transits hit a noise floor at $\approx-10$\,dB (there is a noise bias as we are plotting the positive definite amplitude of the complex data) while beam features in the sidelobes are more clearly discerned in the stack. The middle panel shows all residuals between the stack and the individual transits along with the 2$\sigma$ contours.

The bottom panel compares the observed $1\sigma$ standard error on the mean (i.e. the standard deviation visualized in the middle panel, scaled by a factor $1 / \sqrt{N}$, with $N \sim 100$ in this case) with the fast-cadence estimate of the thermal noise derived from the CHIME real-time pipeline and propagated through the full holography offline pipeline. As we have normalized all transits to be identically $1 + 0j$ on transit, the units of the estimates may be interpreted as a percentage of the on-transit response; in this example, the noise level is $1-2\%$ of the peak. We see that with the exception of some larger excursions in certain pixels, the observed noise follows the trend set by the expected radiometer noise, which sets a lower bound for the observed variation as long as we are not dominated by confusion or other systematics.

For the four dimmer sources presented here, the real-time radiometric noise estimate is generally consistent with what is presented in the bottom panel of Figure \ref{fig:hercules_stack_noise}, i.e. on the order of a percent of the peak or lower. For the two brightest sources, Cygnus A and Cassiopeia A, the radiometric noise is about an order of magnitude lower, at the $<.1\%$ level, and structured variations above the thermal noise limit by a factor of $\sim 2-3$ are discernible. To characterize these features in the stacks we take two different jack-knives of the data, here using Cygnus A. We first consider a seasonal jack-knife; we assign all observations taken between October and March (in any year) to one set, and all other observations taken between April and September to the other set. We will label these as the ``Winter" and ``Summer" partitions, respectively; with this scheme, we assign 56 transits to the summer partition and 49 to the winter partition. We compute the stack over both sets independently, then take their difference. The difference is normalized such that its statistical noise level is equivalent to that of the weighted average of the partitions \citep{CHIMEdetection}. We also consider an alternative but equal (in the size of the sets) partitioning of the data in which transits are assigned randomly to one of two sets. 

Figure \ref{fig:jack} shows a corner plot from the two different jack-knives described above, with the difference between seasonal partitions labeled ``Summer $-$ Winter" and the difference between random partitions labeled ``R0 $-$ R1."  The lower corner plots show noise realizations of the jack-knives; we take the observed variance within each partition and add these in quadrature to estimate the noise level of the jack-knife $\sigma_\text{jack}$, then draw a normal random deviate with mean 0 (i.e. expecting the stacks in each partition to be identical) and standard deviation $\sigma_\text{jack}$. 

It is clear that the jack-knives are not unstructured, zero-mean noise. Along hour angle, the residuals are smallest in the main lobe in both the data-jackknives and noise realizations, indicating that with the current size of the dataset the variability of the beam in this region is comparable to the noise level. However, as we move away from transit to smaller values of the beam response, the fractional residuals dwarf the noise level. Within the frequency-feed subspace, the data-jackknives show vertical striping which is absent from the noise realizations, suggesting systematic variations in frequency which are correlated across the cylinder. Likewise within the hour angle-frequency subspace, there is a clear signature of residual 30\,MHz structure in both data jackknives.

These residuals indicate that for the brightest, highest signal-to-noise sources, we have reached the point of becoming sensitive to and limited by systematic variation in the beam between transits. However, from the comparison of seasonal and random jack-knives, it is not clear that there is systematic variation between observations that is specifically correlated with the ambient temperature. This may have been expected, for example, if daily variation were dominated by expansion / contraction of the focal line with temperature, or if the per-transit normalization scheme were insufficient to account for temperature-dependent gain drifts over the course of the observation. We attempted a further split of the data into four seasonal bins (December - February, March - May, June - August, September - November), yielding an additional six jack-knives. We did not find that the difference of the (coldest) December - February and (warmest) June - August seasons was systematically larger than the other jack-knives, and the amplitudes of the jack-knives (using the variance across voxels as a metric) otherwise do not appear to follow a clear trend with the expected ambient temperature differences of the seasons. The jack-knives with the largest amplitudes tend to involve the spring season, which also has the largest observed variance among its constituent transits. Overall though, the differences in the jackknives saturate at the level of 1\% of the beam value, even in the sidelobes. This is sub-dominant to the $\mathcal{O}(1)$ variation in the sidelobes observed over feed which we noted in Figure \ref{fig:nonredundancy}, and which is implicated in analyses that average over feeds. Some differences could be alleviated by improved cleaning of the data (i.e. masking) on a per-feed and frequency basis for each individual transit.

\subsection{Comparison to solar data} 
\begin{figure*}[ht]
    \centering
    \includegraphics[width=1.00\linewidth]{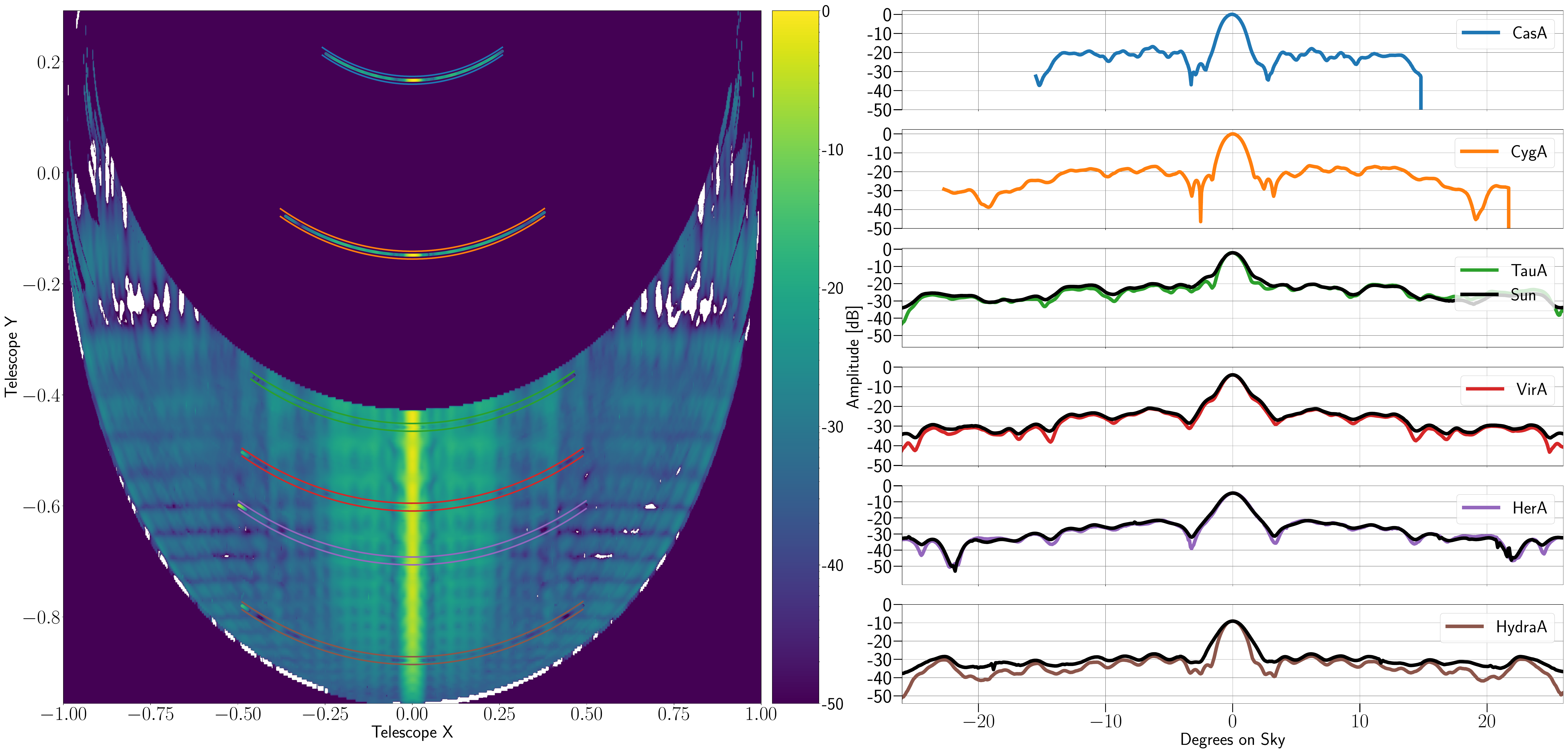}
    \caption{At right: the common-mode primary beam profiles computed from the stacks of six holography sources, in $YY$ polarization at 759\,MHz,  presented in order of descending declination from top to bottom. At left, the solar measurement is plotted in orthographic projection with amplitude mapped to color; holography tracks are overplotted with colored lines outlining the tracks (colored outlines in the left panel correspond to the source with the same color in the 1D panels). Tau A, Vir A, Hercules A, and Hydra A overlap the solar measurement; a slice of the solar data is overplotted in black in the right panels for these overlapping sources.}
    \label{fig:onskyandslices}
\end{figure*}
\label{sec:solarcomp}
\cite{dallassolar} presented measurements of the CHIME primary beam obtained using the Sun as a calibrator source as it passes through a declination range $\pm 23^\circ$ throughout the year. This allows comparisons between holography sources and solar data where they overlap in the southern sky. In this work we present comparisons of solar data to the holography sources Tau A, Virgo A, Hercules A, and Hydra A. 

As described in \cite{dallassolar}, estimates of the primary beam using the Sun were computed by beamforming visibilities corresponding to intra-cylinder baselines under 10\,m long (to avoid resolving out the solar flux) to the location of the Sun. This technique is essentially a coherent average of the visibilities collapsing over the baseline span of the data, such that the resultant measurement is an estimate of the common-mode beam response of CHIME per cylinder. To form a comparable quantity from the holography dataset, we form the holographic estimate of the average \textit{co-polarized} power beam of the baselines used in the solar analysis. In equation form, 

\begin{align}
    \label{eq:co_hol_template}
    \bar{B}^\text{co, co}(\nu, \phi) &= \\ \nonumber &\frac{1}{N_\text{solar}}\sum_{ij \in \text{solar baselines}}V_{i, 26}^\text{co}(\nu, \phi)V_{j, 26}^\text{*co}(\nu, \phi)
\end{align}

where the superscript ``co" indicates either the $XX$ or $YY$ product and $N_\text{solar}$ is the number of solar baselines.

The comparison between solar data and holography data is shown in Figure~\ref{fig:onskyandslices}. The left hand panel shows a beam map of the solar data and holography data together, indicating regions where the sources in this paper overlap. In the holography tracks, the amplitude is taken from the holography data, scaled to the solar data amplitude on meridian, and shows general alignment in features between the data sets. This is also clear in the right hand panels, which shows profiles for a single frequency in $Y$ polarization.

\begin{figure}[ht]
    \centering
    \includegraphics[width=1.0\linewidth]{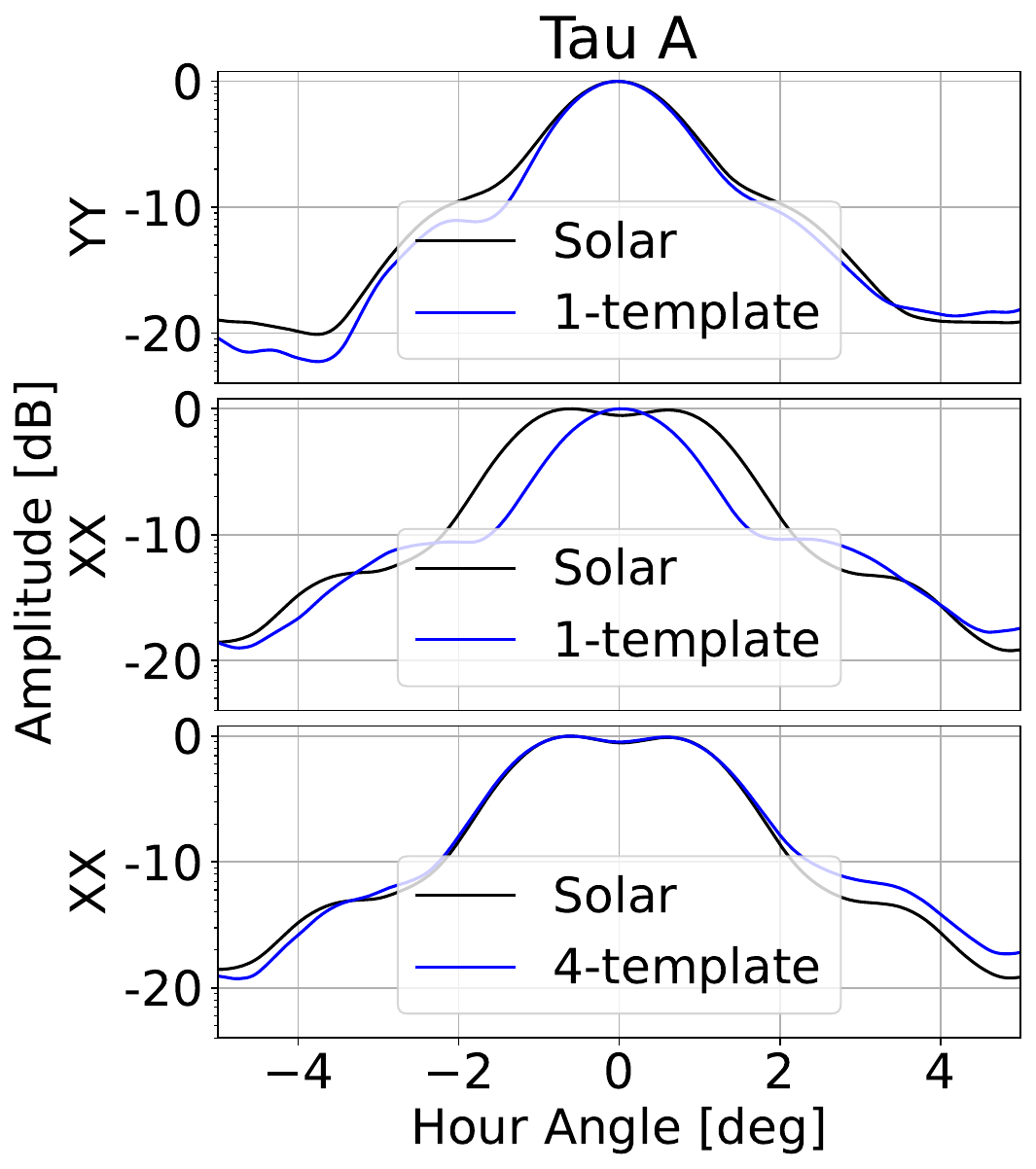}
    \caption{Comparison of the solar beam measurement at the declination of Tau A with the beam profile as measured by holography of Tau A, at 786\,MHz. \textit{Top panel}: In $YY$ polarization, the measurements are comparable to one another. \textit{Middle}: In $XX$ polarization, CHIME's beam sees an excess on either side of the meridian, due to polarization leakage from CHIME $Y$ to CHIME $X$. The holographic measurement here, constructed purely from the co-polarized data, does not include that leakage and appears with an ordinary Gaussian mainlobe. \textit{Bottom}: After including cross-polarized holography in the reconstruction, the ``double bump" feature is successfully captured.}
    \label{fig:doublebump}
\end{figure}

The profiles in Figure \ref{fig:onskyandslices} are generally similar for the solar and holographic response, but a few discrepancies are noteworthy. Far off meridian and particularly at lower declinations, the holography prefers a lower response, as low as $-50$\,dB in certain pixels. Around meridian, the solar measurement tends to prefer a systematically wider beam-width at all declinations, a feature which persists even if the finite size of the Sun on the sky is accounted for by smoothing the holography profiles with a boxcar of about half a degree width. As was discussed in Section \ref{sec:beams}, the beam response of the CHIME $X$ feeds has a systematically wider main lobe than the $Y$ feeds; the discrepant beam-widths in Figure \ref{fig:onskyandslices} are thus suggestive that the solar measurement is probing $X$ signal that has leaked into $Y$. In addition, as shown in the top two panels of Figure \ref{fig:doublebump}, at certain frequencies and in $XX$ polarization only, the main lobe of the solar beam appears significantly distorted from the expected Gaussian shape with an excess in power on either side of meridian. 

These discrepancies between CHIME and co-polarized holography are evidence of the presence of polarization leakage in CHIME; they do not appear in \textit{co-polarized} holography because contaminating leakage does not correlate between like polarization inputs in CHIME and the 26\,m. Schematically, if we write the voltage of an $X$ polarized CHIME input as the sum of $X$ sky signal and signal that has leaked in from $Y$ due to the optics of CHIME:
\begin{equation}
    F_i^{X} \propto E^{X} + E^{Y\rightarrow X}
\end{equation}
then when this voltage is correlated with the voltage of another CHIME input, there will be an overall contribution from the correlation of the leakage terms. However, if the 26\,m receiver does not leak signal in this way, so that
\begin{equation}
   F_{26}^{X} \propto E^{X} , 
\end{equation}
then the leaked $Y$ signal above will decorrelate when forming a co-polarized visibility with the 26\,m's $X (Y)$ input. Instead, we should expect to see the effect of the leakage in the \textit{cross-polarized} holography, i.e. in $V^{X, Y}_{i, 26} = \left<F_i^{X}F_{26}^{Y}\right>$.

Thus, to reconstruct the solar measurement from the holography, we must form a combination of the co- and cross-polarized holography. Generalizing from Equation \ref{eq:co_hol_template}, 
\begin{equation}
    \label{eq:all_hol_template}
    \bar{B}^{pq}(\nu, \phi) = \frac{1}{N_\text{solar}}\sum_{ij \in \text{solar baselines}}V_{i, 26}^{p}(\nu, \phi)V_{j, 26}^{*q}(\nu, \phi)
\end{equation}
where the indices $p, q \in \{\text{co}, \text{cross}\}$. So, for each polarization, we now have four templates $\{\bar{B}^\text{co, co}, \bar{B}^\text{cross, cross}, \bar{B}^\text{co, cross}, \bar{B}^\text{cross, co}, \}$. We smooth each of these templates with a boxcar approximating the angular size of the Sun. We then model the solar measurement as some linear combination of these and solve for the coefficients with a linear least squares fit.

\begin{figure}[ht]
    \centering
    \includegraphics[width=1.0\linewidth]{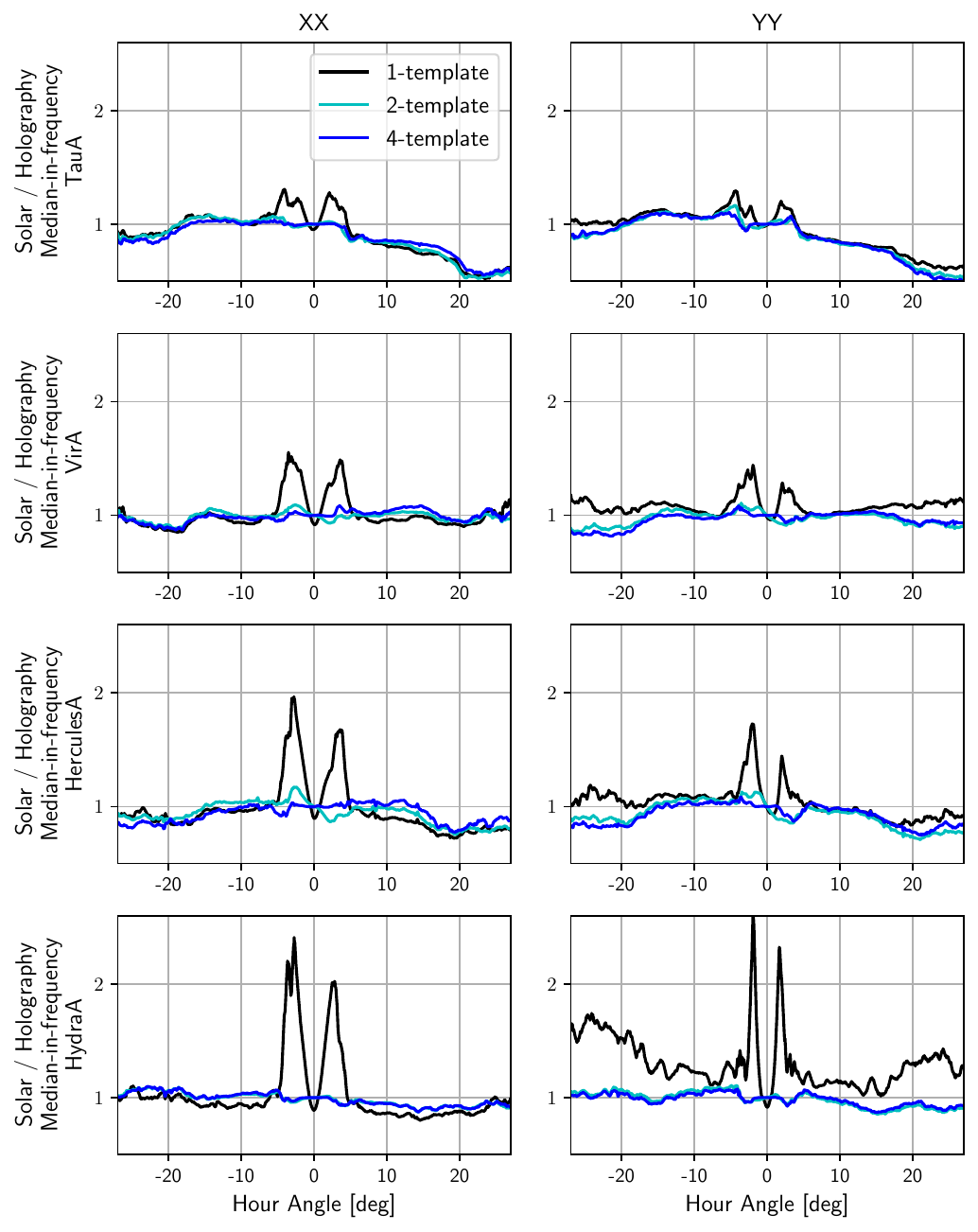}
    \caption{The median over frequency of the ratio of the solar measurement of the primary beam with the 1-, 2-, and 4-template best-fit holographic reconstructions, for each of the four sources overlapping the solar declination range and in each co-polarization.  In general, the inclusion of cross-polarization information in the reconstruction dramatically improves agreement between the two datasets in the main lobe, where beam-width and primary-null level discrepancies due to polarization leakage cause the holography to deviate from the solar measurement by as much as a factor $> 2$ at the declination of Hydra A.}
    \label{fig:med_solar_holo}
\end{figure}

As seen in the bottom panel of Figure \ref{fig:doublebump}, this procedure resolves the distortion in the main lobe at certain frequencies where the polarization leakage is particularly severe. We summarize the results of this comparison procedure for the 3 other southern holographic sources in Figure \ref{fig:med_solar_holo}, which show medians (over frequency) of the ratio of the solar response and the holographic reconstruction as described above, with the naive 1-template fit (i.e. fitting an overall amplitude) in black, a 2-template fit in cyan (which does not include the products of co- and cross-polarized holography, only co- with co- and cross- with cross-) and the full 4-template fit in blue. The inclusion of cross-polarized holography removes the large excursions on either side of transit for all sources. For Hydra A, which is biased low compared to the solar measurement across the entire East-West range seen in Figure \ref{fig:onskyandslices}, the inclusion of cross-polarized information brings the holography into close agreement with the solar measurement on average through the band. For these sources, though, there remain polarization and declination-dependent deviations as large as 45\% (Tau A) in the far sidelobes. 

There is little improvement in the reconstruction when including all 4 templates compared to only 2, suggesting that the terms $\bar{B}^\text{co, cr}, \bar{B}^\text{cr, co}$ are relatively unimportant and do not explain the remaining residuals with the solar measurement. These terms are expected to contribute only if the 26\,m has its own polarization leakage. As a result, we conclude that any leakage in the 26\,m system is subdominant to the source of the discrepancies observed in Figure \ref{fig:med_solar_holo}. A possible source of this discrepancy is differential gain of the 26\,m receiver; i.e. even if the 26\,m Jones matrix is diagonal, the two polarization channels may not have equivalent gain. In this case, the 2-template holographic reconstruction will differ from the CHIME data by an additional term which depends both on parallactic angle and the spatial-dependence of the CHIME Jones matrix. 

\subsection{Point-source subtraction}

Section~4.1 of \cite{CHIMEdetection} describes a stage of the CHIME data processing pipeline that performs a targeted removal of the four brightest point sources (Cygnus~A, Cassiopeia~A, Taurus~A, and Virgo~A) from the CHIME data by fitting and subtracting the following ``simple model'' from the visibilities:
\begin{align}
    \label{eq:model_bright_psrc}
    \mathcal{V}_{\rm psrc}(\vec{u}, \nu, t) = \sum_{s=1}^{4} a_{s}(\nu, t) \ e^{i 2 \pi \vec{u} \cdot \vec{\hat{n}}_s(t)} \ .
\end{align}
Here $\vec{u}$ is the baseline vector, $\nu$ is the frequency, $t$ is the time,  $\vec{\hat{n}}_s(t)$ the direction of the source on the sky at time $t$. The source amplitude $a_{s}(\nu, t)$ is meant to encode the spectral flux density of source~$s$ modulated by the primary beam pattern of the instrument at the source location.  The amplitudes are estimated for each frequency and time by performing a weighted linear regression over the inter-cylinder baselines.  After performing a 2D smoothing of the best-fit amplitudes in frequency and time, the resulting model is subtracted from the visibilities.

The model given by Equation~\ref{eq:model_bright_psrc} assumes that the source amplitudes are constant as a function of baseline, which is true if the angular extent of the source is much less than $1 / \vec{u}_{\rm max} \approx 10 \ \mbox{arcmin}$, if residual complex gain variations are negligible or common to all feeds, and if the primary beam pattern is the same for all feeds.  However, the holography observations indicate significant feed-to-feed variations in the primary beam pattern, which violate the last assumption.  The feed-to-feed variation was explored in Section~\ref{sec:beams}.  In this section, we attempt to account for the feed-to-feed variation in the point source subtraction algorithm by incorporating the holographic measurements into our model for the signal from the four brightest sources.

The new ``holography-based model'' for the visibilities is given by
\begin{align}
    \label{eq:model_bright_psrc_holo}
    \mathcal{V}_{\rm psrc}^{\rm hol}(\vec{u}, \nu, t) = \sum_{s=1}^{4} \sum_{p} \sum_{q} a_{s}^{pq}(\nu) B_{s}^{pq}(\vec{u}, \nu, t) \nonumber \\
     \times e^{i 2 \pi \vec{u} \cdot \vec{\hat{n}}_s(t)} \ . 
\end{align}
Here again $p, q \in \{\mbox{co-polar}, \mbox{cross-polar}\}$ and $B_{s}^{pq}$ is the now \textit{baseline-dependent} (note the absence of the overhead bar) set of templates constructed from the holographic observations of source $s$ as follows
\begin{align}
    \label{eq:effective_baseline_beam}
    B_{s}^{pq}(\vec{u}, \nu, t) = \frac{1}{N_{b}}\sum_{ij \in \vec{u}} V^{s, p}_{i, 26}(\nu, \phi) V^{*s, q}_{j, 26}(\nu, \phi)
\end{align}
where the summation runs over the $N_{b}$ redundant pairs of feeds separated by baseline vector $\vec{u}$.  The 16 coefficients $a_{s}^{pq}(\nu)$ (corresponding to the 4 sources and 4 polarisation pairs) are estimated for each frequency by performing a weighted linear regression over the inter-cylinder baselines and all time samples spanning a full sidereal day.  Since the holographic measurements describe the time dependence of the source amplitudes, the total number of free parameters that must be solved for is greatly reduced compared to the simple method.

\begin{figure*}[ht]
    \centering
    \includegraphics[width=1.0\linewidth]{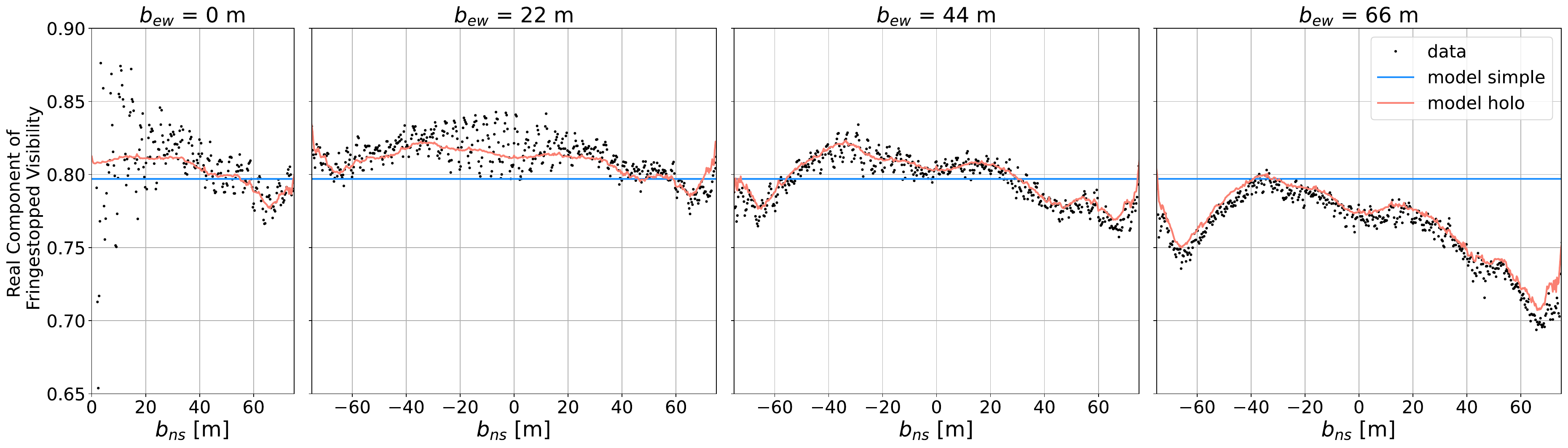}
    \caption{CHIME visibilities (XX polarisation) as a function of baseline distance at a frequency of 612.5 MHz for a single 10 second time integration acquired on October 24, 2019.  At this time, the visibilities are dominated by the signal from Cygnus~A, which is in the main lobe of the primary beam at an hour angle of $1^{\circ}$.  The visibilities have been fringestopped to the location of Cygnus~A and normalized so that they are equal to 1 at the transit of Cygnus~A.  Each column corresponds to a different east-west baseline distance with the x-axis denoting the north-south baseline distance.  The redundant copies of each baseline have been averaged.  The data is shown in black, the best-fit, simple model given by Equation~\ref{eq:model_bright_psrc} is shown in blue, and the best-fit, holography-based model given by Equation~\ref{eq:model_bright_psrc_holo} is shown in red.  The visibilities exhibit $\sim$10\% variations across baselines, which are not captured by a simple point-source model, but are well described by a model consisting of a point source modulated by the feed-dependent primary beam pattern inferred from holographic observations of this source.  The fine structure present in the visibilities as a function of north-south baseline distance is due to fainter sources within the CHIME field of view that are not being modelled during the fit.}
    \label{fig:vis_psrc_sub}.
\end{figure*}

Figure~\ref{fig:vis_psrc_sub} compares the best-fit, simple model and the best-fit, holography-based model to the measured visibilities as a function of baseline for a single frequency channel and time integration.  The dominant contribution to the visibilities at this time is the signal from Cygnus~A, which is at an hour angle of only $1^{\circ}$.  The visibilities vary by approximately $10\%$ as a function of baseline.  The simple point source model is unable to describe this variation by definition.  However, the holographic observations of Cygnus~A accurately predict this variation, indicating that it is caused by a change in the effective primary beam pattern with baseline.  At this specific hour angle of $1^{\circ}$, the variation is mostly driven by wander in the centroid of the primary beam with the position of the feed along the cylinder, which is illustrated in Figure~\ref{fig:centroids} and whose effect is also highlighted in Figure~\ref{fig:nonredundancy}. Larger baseline separations exhibit larger deviations from the mean behavior, which is described by the simple model, because fewer redundant baselines are being averaged.  In addition, those baselines are sampling the centroid wander on different cylinders and/or disparate positions along the cylinder.  After subtracting the holography-based model, the residuals are in general less than $1\%$ of the flux of the source.

\begin{figure}[ht]
    \centering
    \includegraphics[width=1.0\linewidth]{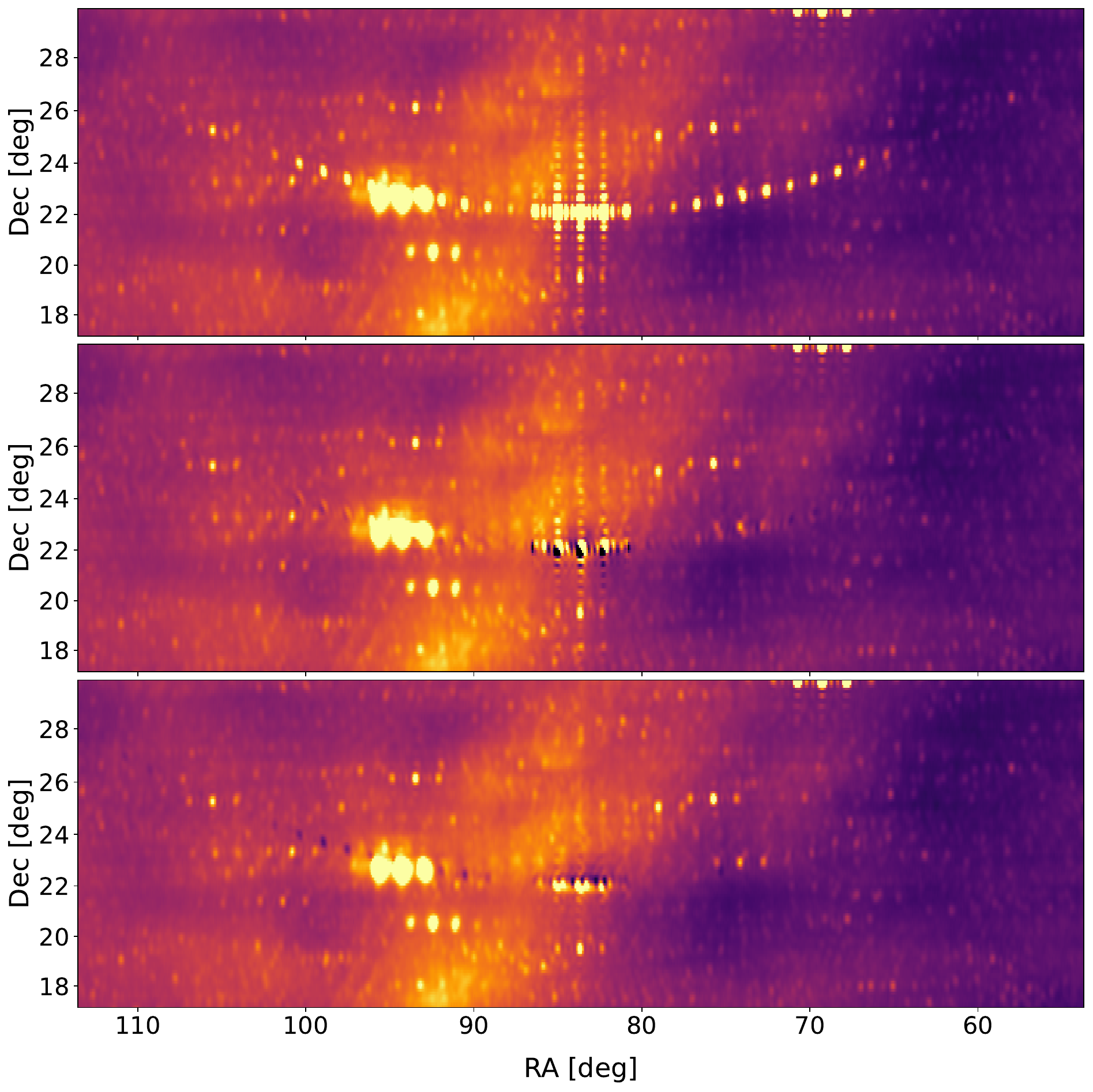}
    \caption{Map of the sky at 612.5 MHz in the region around Taurus~A before point source subtraction (top), after subtracting a simple model for Taurus~A given by Equation~\ref{eq:model_bright_psrc} (middle), and after subtracting a model for Taurus~A based on the holography measurements and given by Equation~\ref{eq:model_bright_psrc_holo} (bottom).  The map was constructed from the XX-polarisation CHIME visibilities acquired on October 24, 2019. The color scale spans [$-0.5\%, \ 1.0\%]$ of the spectral flux density of Taurus~A.}
    \label{fig:ringmap_psrc_sub}
\end{figure}

Figure~\ref{fig:ringmap_psrc_sub} provides an alternative qualification of the effectiveness of the source subtraction by constructing a map from the visibilities focused on the region around Taurus~A before and after applying the different algorithms.  The map making procedure is described in \cite{CHIMEoverview}.  The holography-based subtraction results in a significant improvement in the quality of the residual map compared to the simple method.  This is because it accounts for feed-to-feed variations in the primary beam, but also because it fully constrains the time dependence of the source amplitudes, greatly reducing the number of degrees of freedom and preventing the algorithm from subtracting the unmodelled, background sky.  Residual flux from the source is observed to be $\lesssim 1\%$ of the peak flux.  However, the holography-based method clearly oversubtracts the signal when the source is in the far side lobes at positive hour angle (most obvious between $90^{\circ} \lesssim \mbox{R.A.} \lesssim 100^{\circ}$).  This is consistent with the large fractional discrepancy between the holography and solar-based reconstruction of the far side lobes at positive hour angles at the declination of Taurus~A that is displayed in the top row of Figure~\ref{fig:med_solar_holo}.  This is suggestive of a slow drift in the normalization of the holography observations with respect to the CHIME data.  The origin of this drift is still under investigation. 

Future improvements to the source subtraction algorithm will involve fitting for an overall normalization of the holgraphy-based model that is slowly varying with time in order to capture the unexplained drift in the holographic measurements that are highlighted in Figure~\ref{fig:med_solar_holo}.  In addition, models for the extended emission of each source will also be incorporated into the algorithm.

\section{Conclusion}
\label{sec:conclusion}
In this work we presented an overview of the holographic beam mapping technique adapted for and used to measure the beam pattern of CHIME. This measurement provides us with a rich dataset, allowing us to measure the beam response of all CHIME feeds in amplitude and phase; and providing constraints on the beam-phase and variation between feed elements that are uniquely accessible to the holographic technique among CHIME's beam-mapping methods. The relatively small confusion noise limit of the measurement enables highly repeatable measurements of the beam sidelobes across a wide range of declinations, albeit with sparse sampling. 

We have detailed an offline analysis pipeline accounting for known systematic effects of the measurement and enabling combination of repeated observations of each source to mitigate thermal noise. This pipeline has been demonstrated on a small portion of the available data, for a subset of six of the brightest sources available for holography. 

We summarized these beam measurements as a function of frequency and feed vs position on the sky. These full 2D visualizations illustrate the frequency and feed dependence of structure in the sidelobes while most of the observed variation in the region of the main lobe is more concisely described by the best-fit centroids and beam-widths. We use the phase information provided by holography to sketch out some basic features of the CHIME aperture, and the possible interaction between the cylinders. 

We have cross-checked the resulting beam measurements against independent CHIME data, including the solar beam measurement and visibility timestreams. Initial discrepancies in the main lobe region, particularly the absence, in co-polarized holography, of the ``double-lobe" distortion of Figure \ref{fig:doublebump} provided an early indication of the presence of polarization leakage in CHIME. When this leakage is accounted for by including cross-polarized holography in attempts to reconstruct the independent CHIME measurements, we find much-improved agreement in the main lobe region. However, as illustrated in Figure \ref{fig:med_solar_holo} and the asymmetric subtraction of Tau A in Figure \ref{fig:ringmap_psrc_sub}, there remain discrepancies between holography and independent CHIME data in the far sidelobes which are declination-dependent and as large as 40\% in some directions. 

We have demonstrated a few key respects in which even the limited sample of observations presented here has provided invaluable insights on, and paths to further investigation of, the properties of CHIME. As we proceed with the full data set, we anticipate that holography's uniquely detailed window into the East-West sidelobes for these sources will provide important input to our beam modeling efforts in accurately capturing the behavior of the instrument in these regions on the sky. This should in turn facilitate an improved characterization of the signatures of foregrounds in the sidelobes as CHIME continues to pursue the \tcm power spectrum.

\section*{Acknowledgements}

We thank Meiling Deng for her substantial contributions to the
instrument design and analysis of CHIME.

We thank the Dominion Radio Astrophysical Observatory, operated by the National
Research Council Canada, for gracious hospitality and expertise. The DRAO is
situated on the traditional, ancestral, and unceded territory of the Syilx
Okanagan people. We are fortunate to live and work on these lands.

We acknowledge that the holographic measurements presented in this work depended on DRAO's John A. Galt 26\,m telescope. We thank the DRAO for their generosity in allowing CHIME to make extensive use of the Galt telescope. We extend special thanks to Tim Robishaw and Andrew Gray for coordinating observing time on the Galt telescope, to Benoit Robert, Rob Messing, and Mike Smith for supporting the operation of the Galt telescope, and to Andre Johnson for designing the customized CHIME receiver for the Galt telescope.

CHIME is funded by grants from the Canada Foundation for Innovation (CFI) 2012
Leading Edge Fund (Project 31170), the CFI 2015 Innovation Fund (Project 33213),
and by contributions from the provinces of British Columbia, Québec, and
Ontario. Long-term data storage and computational support for analysis is
provided by WestGrid, SciNet and the Digital Research Alliance of Canada, and we
thank their staff for flexibility and technical expertise that has been
essential to this work.

Additional support was provided by the University of British Columbia, McGill
University, and the University of Toronto. CHIME also benefits from NSERC
Discovery Grants to several researchers, funding from the Canadian Institute for
Advanced Research (CIFAR), from Canada Research Chairs, from the FRQNT Centre de
Recherche en Astrophysique du Québec (CRAQ) and from the Dunlap Institute for
Astronomy and Astrophysics at the University of Toronto, which is funded through
an endowment established by the David Dunlap family. This material is partly
based on work supported by the NSF through grants (2008031) (2006911) and
(2006548) and (2018490) and by the Perimeter Institute for Theoretical Physics, which in turn
is supported by the Government of Canada through Industry Canada and by the
Province of Ontario through the Ministry of Research and Innovation.

K.W. Masui holds and acknowledges the support of the Adam J. Burgasser Chair in Astrophysics. 

\software{
bitshuffle \citep{2015bitshuffle},
caput \citep{caput},
ch\_pipeline \citep{ch_pipeline},
cora \citep{cora},
Cython \citep{Cython},
draco \citep{draco},
driftscan \citep{driftscan},
hankl \citep{karamanis2021},
h5py \citep{h5py},
HDF5 \citep{HDF5},
HEALPix \citep{healpix},
healpy \citep{healpy},
Matplotlib \citep{Matplotlib},
mpi4py \citep{mpi4py},
NumPy \citep{NumPy},
OpenMPI \citep{OpenMPI},
pandas \citep{pandas,pandas_paper},
peewee \citep{Peewee},
SciPy \citep{SciPy},
Skyfield \citep{Skyfield},
}

\appendix

\section{CHIME PFB and Decorrelation correction}
\label{sec:decorr_appendix}
As described in \cite{CHIMEoverview}, the CHIME correlator follows an FX design where the analog signals from the CHIME antennas are first fed into an FPGA-based system (the ``F-Engine") which samples, digitizes and channelizes the timestreams into a 400-800\,MHz band subdivided into 1024 channels, each of width 390.625\,kHz. 

The channelization step is implemented in FPGA firmware, as a polyphase filter bank (PFB) \citep{parsonsfpga}, which aims to ensure that the frequency response within a given channel is flat and dies rapidly outside the channel. The PFB performs the following steps on the input digitized signal: 
\begin{enumerate}
    \item Accumulate a finite number of digitized samples in time, $x[t_{i}]$, organized in $M$ chunks each of $N$ samples. 
    \item Apply a window function $w[t_{i}]$ of length $MN$-samples to the accumulated data.
    \item Perform a Fast Fourier Transform on the windowed data.
    \item Select only every $M$-th output channel. 
\end{enumerate}

For the PFB implemented in CHIME, $M = 4, N = 2048$ (and the underlying timestreams are sampled at 800\,MSPS) and the window function $w$ is a sinc-Hamming window
\begin{equation}
    w[t_{i}] = \sinc\left({\frac{t_{i}}{N} - \frac{M}{2}}\right)\left(0.54 - 0.46\cos{\frac{2\pi t_{i}}{NM -1}}\right)
\end{equation}

The decorrelation which affects holography is a consequence of the fact that we are channelizing and correlating frames of finite length, so that when one timestream is delayed relative to another, some of the correlated information is not captured within the same frame. When correlating CHIME-only signals, the relevant delays between signals consist of only the expected geometric delays central to interferometry and are too small to lead to any appreciable signal loss. In the case of holography, however, the overall delay, due to the extra lengths of cables connecting the 26\,m to the CHIME correlator, becomes a significant fraction of the $2.56\mu$s integration window and the strength of the decorrelation becomes significant. 

Given the implementation of the PFB, it is possible to calculate (and thus correct for) the amount of signal loss due to decorrelation when the input signals are delayed by time $\tau$. Note that this approach is not ``recovering signal," but rather compensating for the fact that as the decorrelation varies continuously with $\tau$ (and thus with time during an observation as the geometric delay associated with the transiting source changes), the primary effect we are sensitive to in the holography is a distortion of the beam shape, which we attempt to estimate and correct for here. The output of the PFB can be written as 

\begin{equation}
\label{eq:pfb_x}
    \tilde{x}[q, f] = \int d\nu e^{2\pi i \nu qN\Delta t}\tilde{w}\left(\nu -\nu_{f}\right)\tilde{x}(\nu)
\end{equation}
where $q$ indexes the $2.56\mu$s output integration frames, $f$ is a channelized frequency with $\nu$ the underlying continuous frequency of the signal $x$, the overhead tildes denote Fourier transforms, $\nu_{f} = f / (N\Delta t)$ and $\Delta t$ is the cadence in time of the digitized samples. 

The decorrelation results from taking a correlation of Eq. \ref{eq:pfb_x} with an identical copy of itself but delayed by time $\tau$. Taking this correlation and assuming that the underlying signal $x$ has stationary statistics with power spectrum $V(\nu)$ (i.e. the visibility spectrum we are trying to measure) we find

\begin{equation}
    \left<\tilde{x}\tilde{x}_{\tau}\right> = V(\nu_{f} + \nu_{n})e^{2\pi i(\nu_{f} + \nu_{n})\tau}\int_{-\nu_{n} /2}^{\nu_{n} /2}d\nu e^{2\pi i\nu\tau}|\tilde{w}(\nu)|^2
\end{equation}

where $\nu_{n}$ is the Nyquist frequency associated with the $2.56\mu$s integration cadence (i.e., the width of the output channels, 390\,kHz). The term outside the integral is essentially the ideal visibility we wish to measure, but it is modulated by the value, at delay $\tau$, of the Fourier transform of the square of the PFB window function. When $\tau$ is large, this term suppresses the measured visibility. Hence we can define the decorrelation ratio as

\begin{equation}
    D(\tau) = \frac{\int_{-\nu_{n} / 2}^{\nu_{n} / 2}d\nu e^{2\pi i\nu\tau|\tilde{w}(\nu)|^2}}{\int_{-\nu_{n}/2}^{\nu_{n}/2}|\tilde{w}(\nu)|^2}
\end{equation}

\begin{figure}
    \centering
    \includegraphics[width=.5\linewidth]{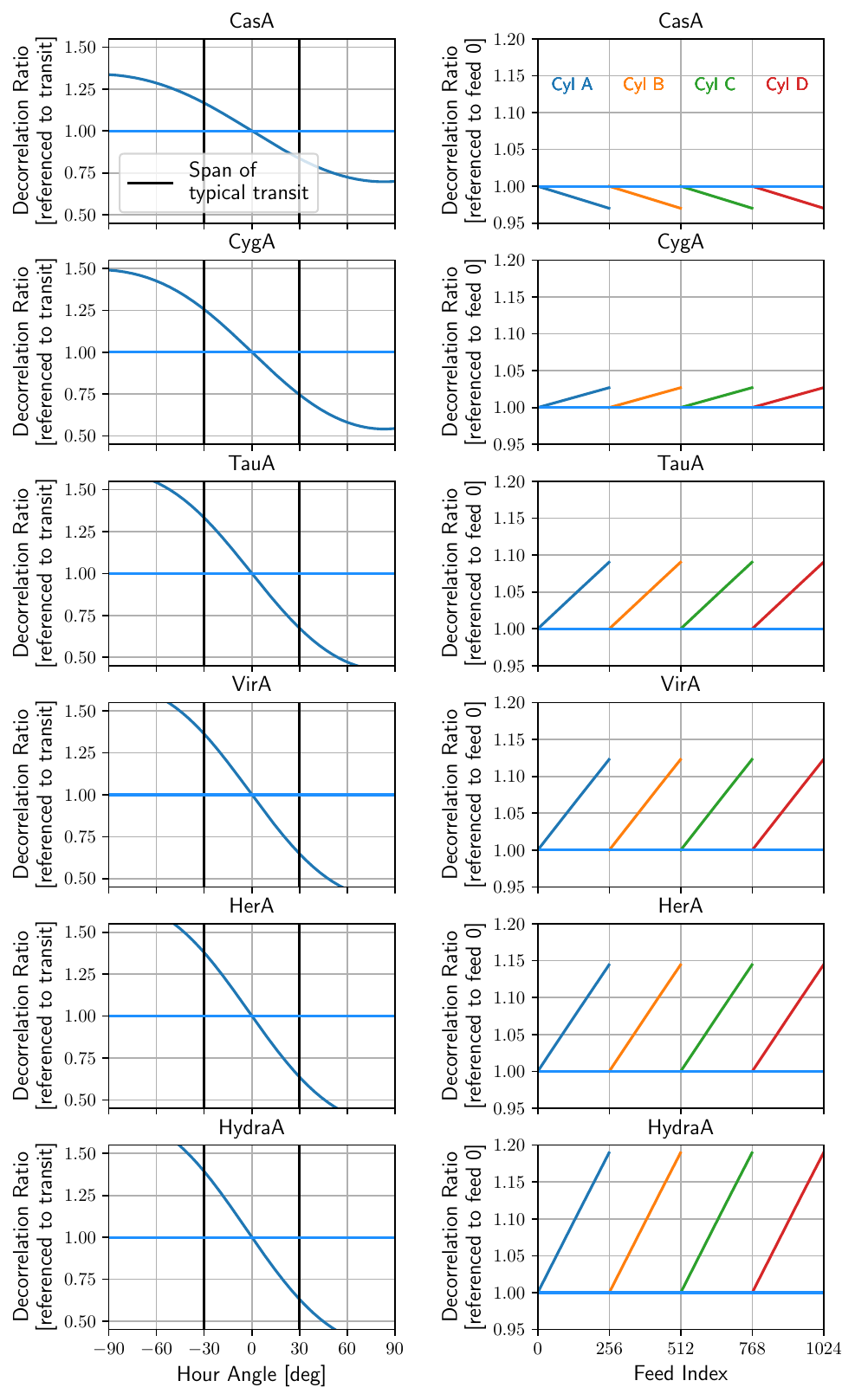}
    \caption{The effect of the decorrelation between CHIME and the Galt telescope on the amplitude of the holography measurement. Each row depicts a different source, in order from highest declination at the top to lowest at the bottom. In the left column we show the amplitude of the decorrelation as a function of hour angle, referenced to 1 at transit. The black vertical lines in these panels indicate the span ($\pm30^\circ$) of the hour angle grid used for all data presented in this paper; within this range the effect is approximately linear but becomes nonlinear in the very far sidelobes. We do not consider this regime here but use the full result of the calculation when correcting the data. In the right column, we show the dependence of the decorrelation strength on transit as a function of feed along the cylinders (color-coded), referenced to the first feed. Each cylinder behaves in the same way; the slope becomes more severe at larger zenith angles and flips sign across zenith.}
    \label{fig:decorr-dependence}
\end{figure}

For a given $\tau$, this expression can be calculated numerically. In Figure \ref{fig:decorr-dependence}, we show the result as a function of hour angle and feed index for the six sources presented here. Noting that we have an unconstrained overall amplitude scale in the holography, we reference the result of the calculation to unity at transit (HA = 0) in the hour angle plots and to feed 0 on the cylinders in the feed index plots.

\pagebreak
\section{Figures for other sources}
In this Appendix we present versions of Figures \ref{fig:freqwaterfall}, \ref{fig:feedwaterfall} for each of the 5 other sources included in this paper: Cas A, Tau A, Virgo A, Hercules A, and Hydra A. Figure \ref{fig:decorr-dependence} shows the calculated effect of the decorrelation discussed in Section \ref{subsec:fringeregriddecorr}.

\begin{figure*}[ht]
    \centering
\includegraphics{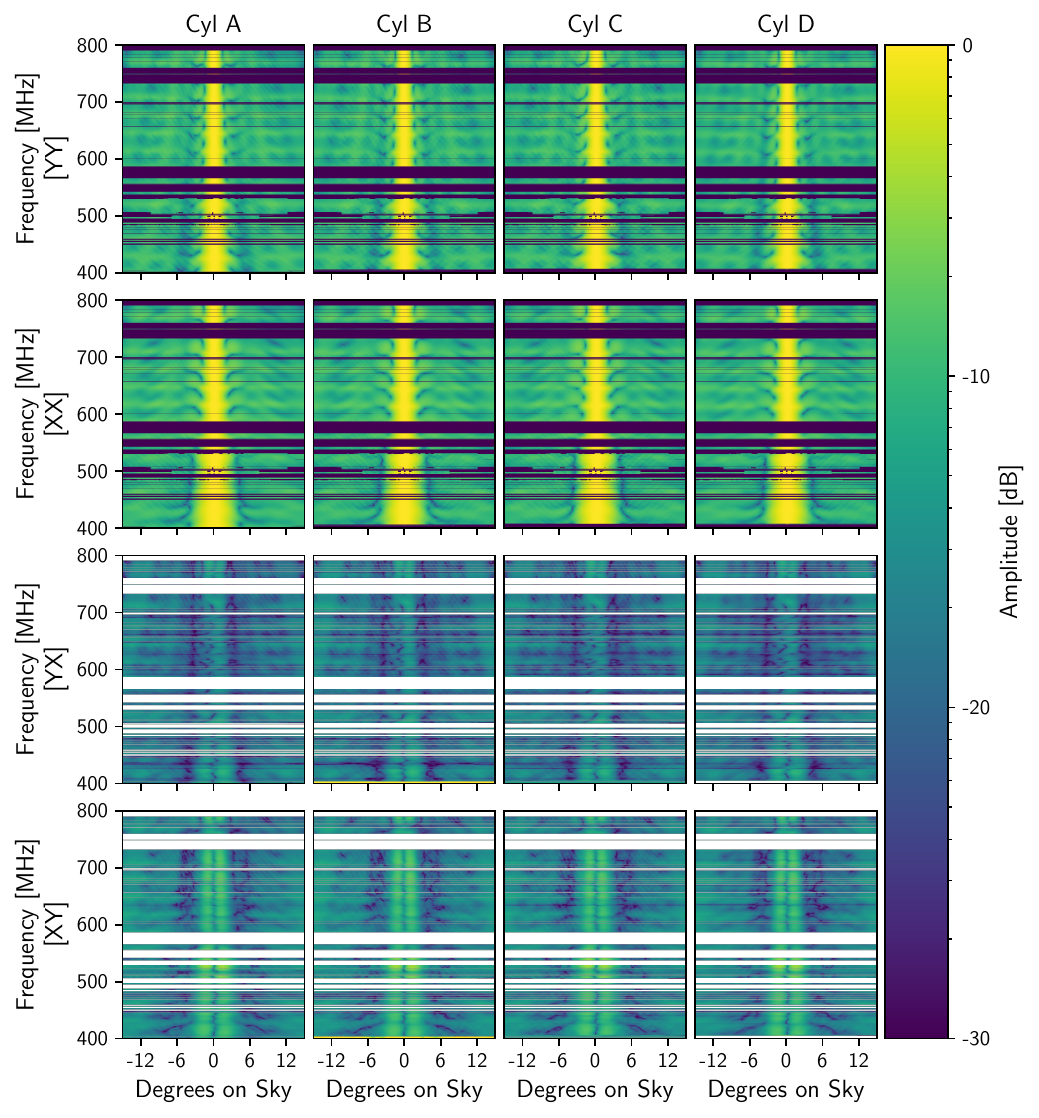}
    \caption{Waterfall plots (in frequency vs sky degrees) of the stacked Cas A data.}
    \label{fig:freqwaterfallCas}
\end{figure*}

\begin{figure*}[t]
    \centering
    \includegraphics{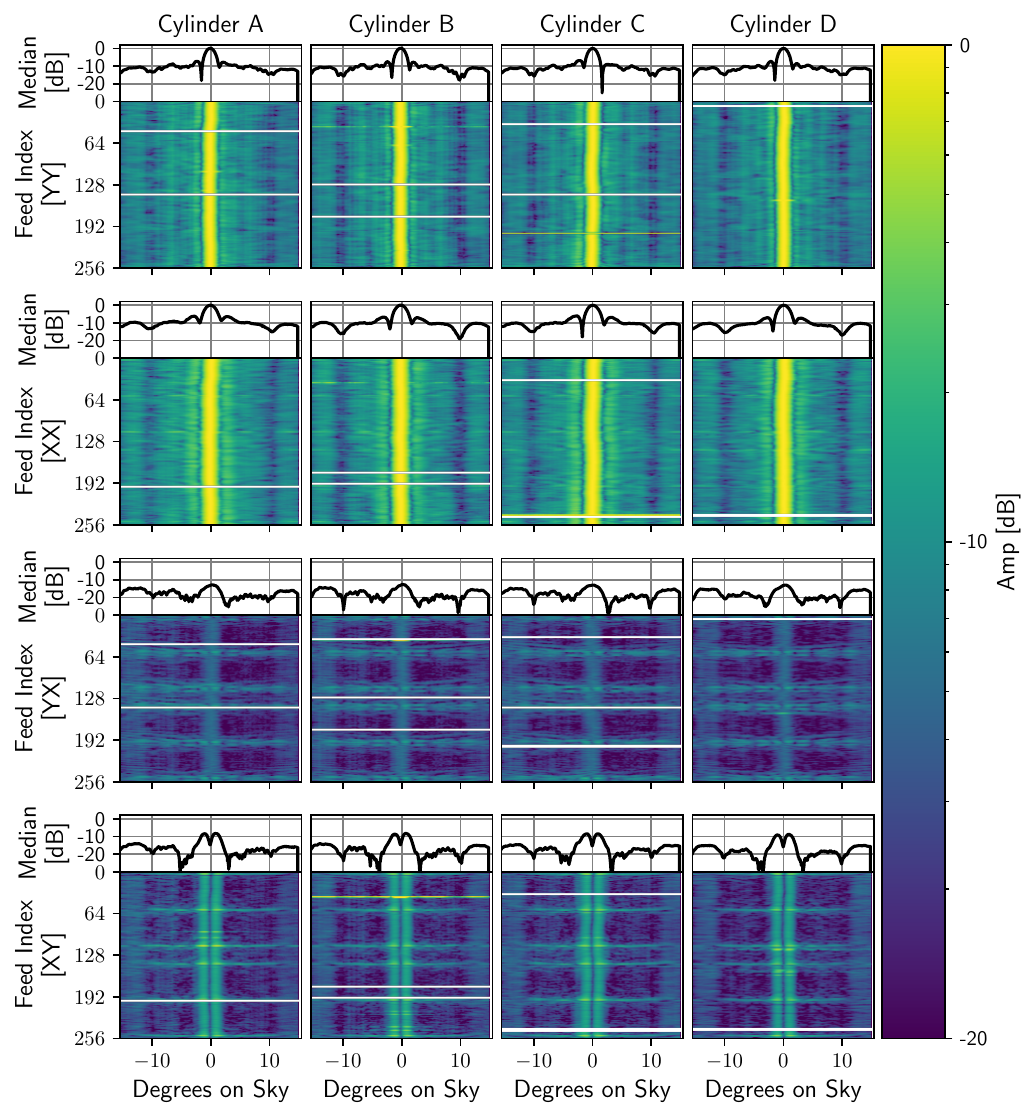}
    \caption{Waterfall plots (in feed index vs. sky degrees) of the stacked Cas A data. For each cylinder data is taken from 717 MHz. The top two rows show co-pol data while the bottom two rows show cross-pol. The one dimensional slices above each image panel show the median along the respective cylinders.}
    \label{fig:feedwaterfallCas}
\end{figure*}

\begin{figure*}[ht]
    \centering
\includegraphics{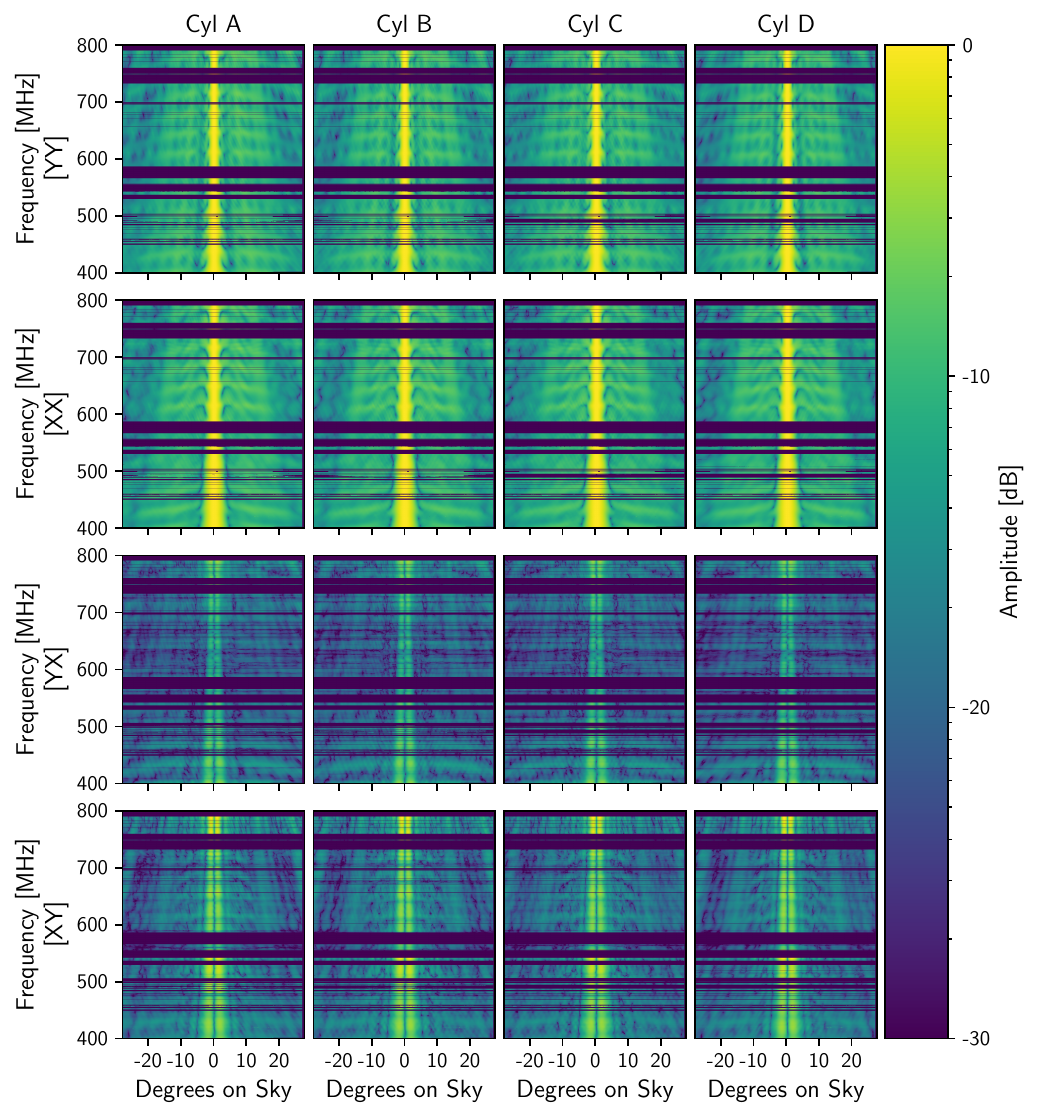}
    \caption{Waterfall plots (in frequency vs sky degrees) of the stacked Tau A data.}
    \label{fig:freqwaterfallTau}
\end{figure*}

\begin{figure*}[t]
    \centering
    \includegraphics{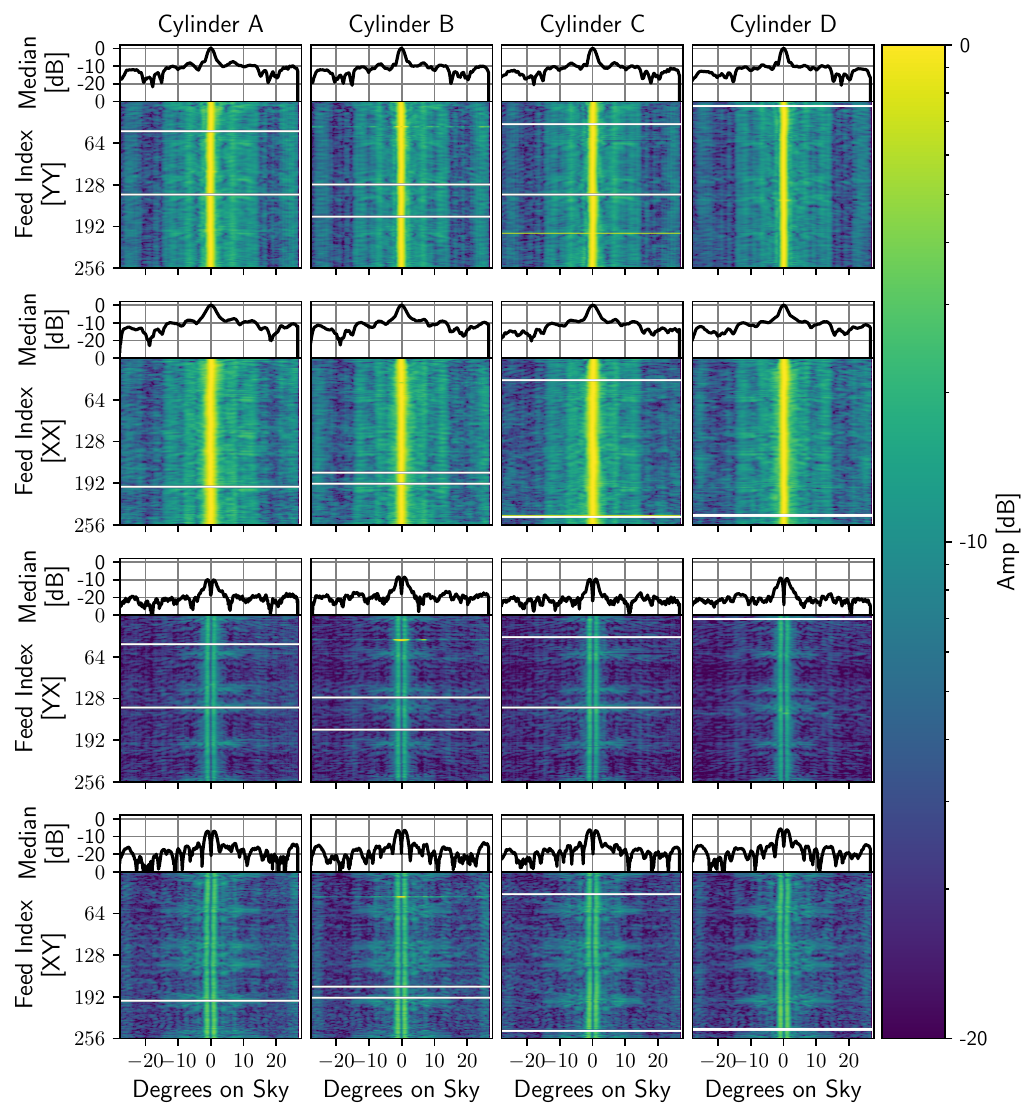}
    \caption{Waterfall plots (in feed index vs. sky degrees) of the stacked Tau A data. For each cylinder data is taken from 717 MHz. The top two rows show co-pol data while the bottom two rows show cross-pol. The one dimensional slices above each image panel show the median along the respective cylinders.}
    \label{fig:feedwaterfallTau}
\end{figure*}

\begin{figure*}[ht]
    \centering
\includegraphics{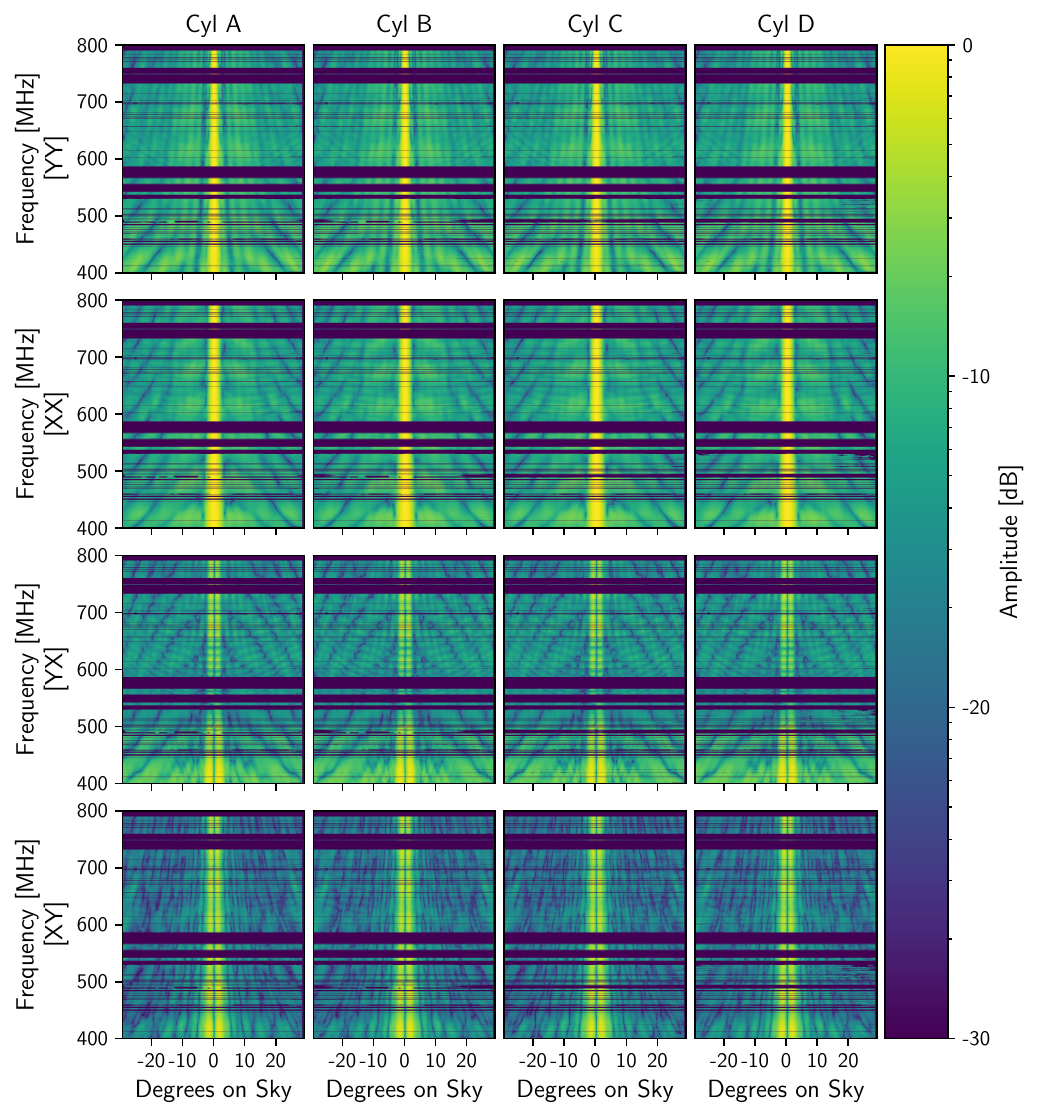}
    \caption{Waterfall plots (in frequency vs sky degrees) of the stacked Vir A data.}
    \label{fig:freqwaterfallVir}
\end{figure*}

\begin{figure*}[t]
    \centering
    \includegraphics{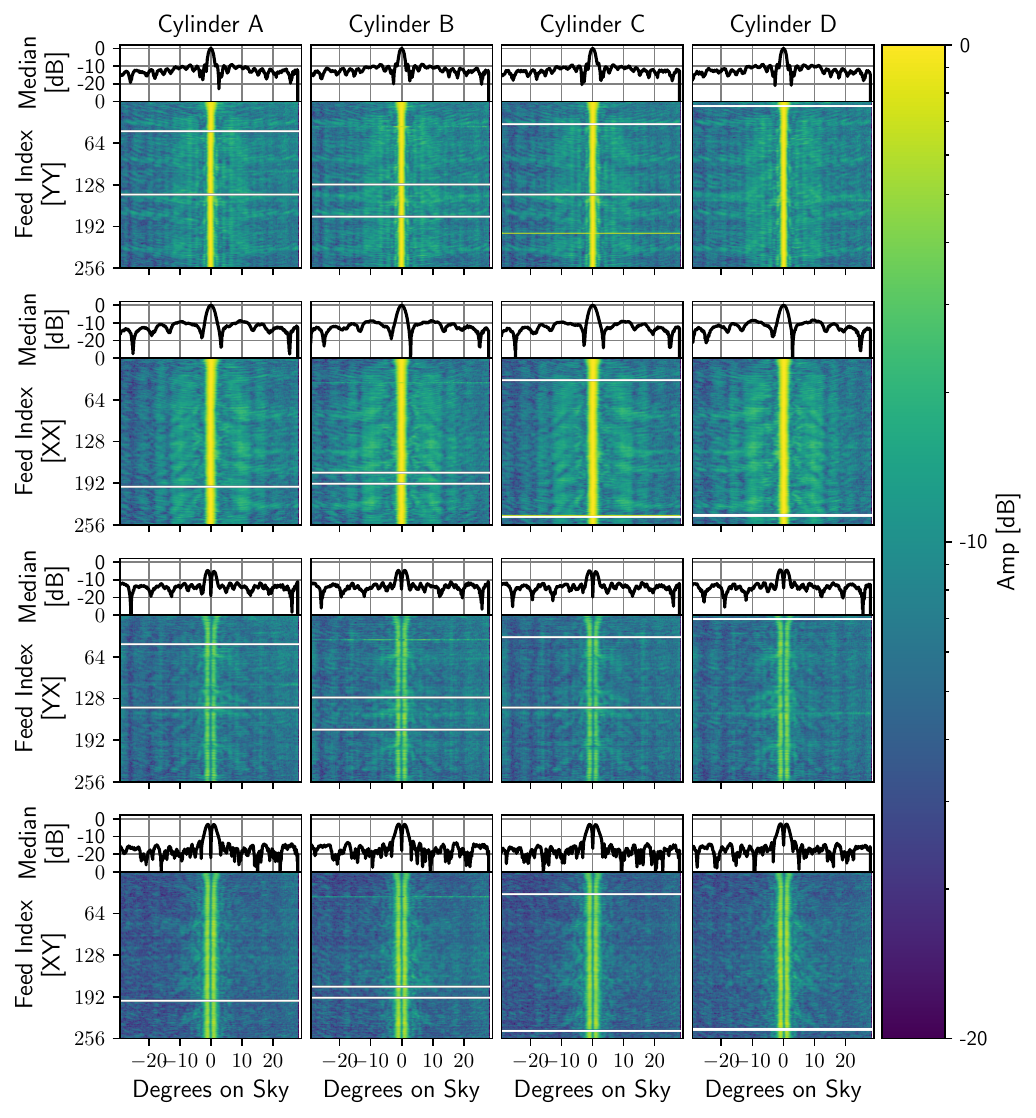}
    \caption{Waterfall plots (in feed index vs. sky degrees) of the stacked Vir A data. For each cylinder data is taken from 717 MHz. The top two rows show co-pol data while the bottom two rows show cross-pol. The one dimensional slices above each image panel show the median along the respective cylinders.}
    \label{fig:feedwaterfallVir}
\end{figure*}

\begin{figure*}[ht]
    \centering
\includegraphics{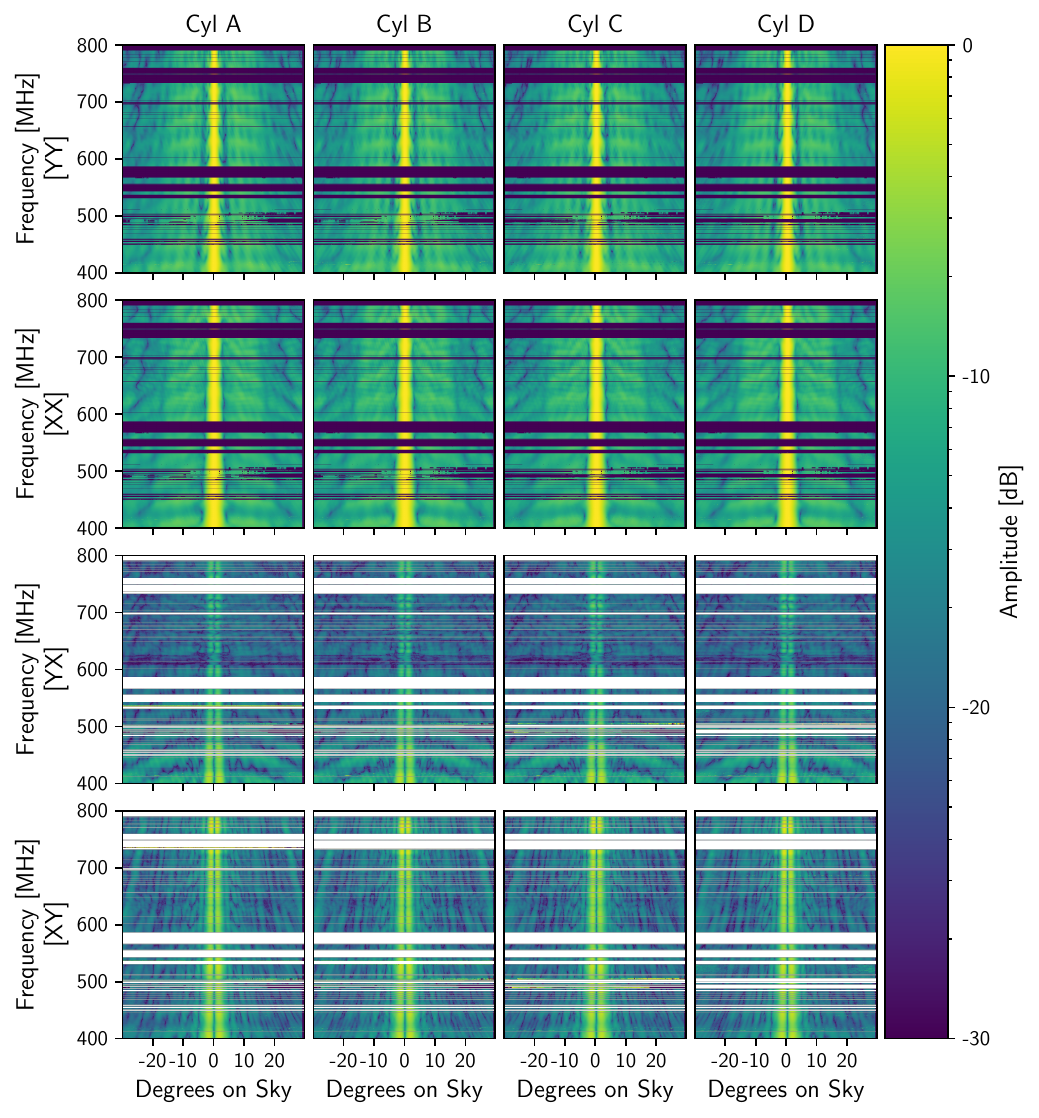}
    \caption{Waterfall plots (in frequency vs sky degrees) of the stacked Her A data.}
    \label{fig:freqwaterfallHer}
\end{figure*}

\begin{figure*}[t]
    \centering
    \includegraphics{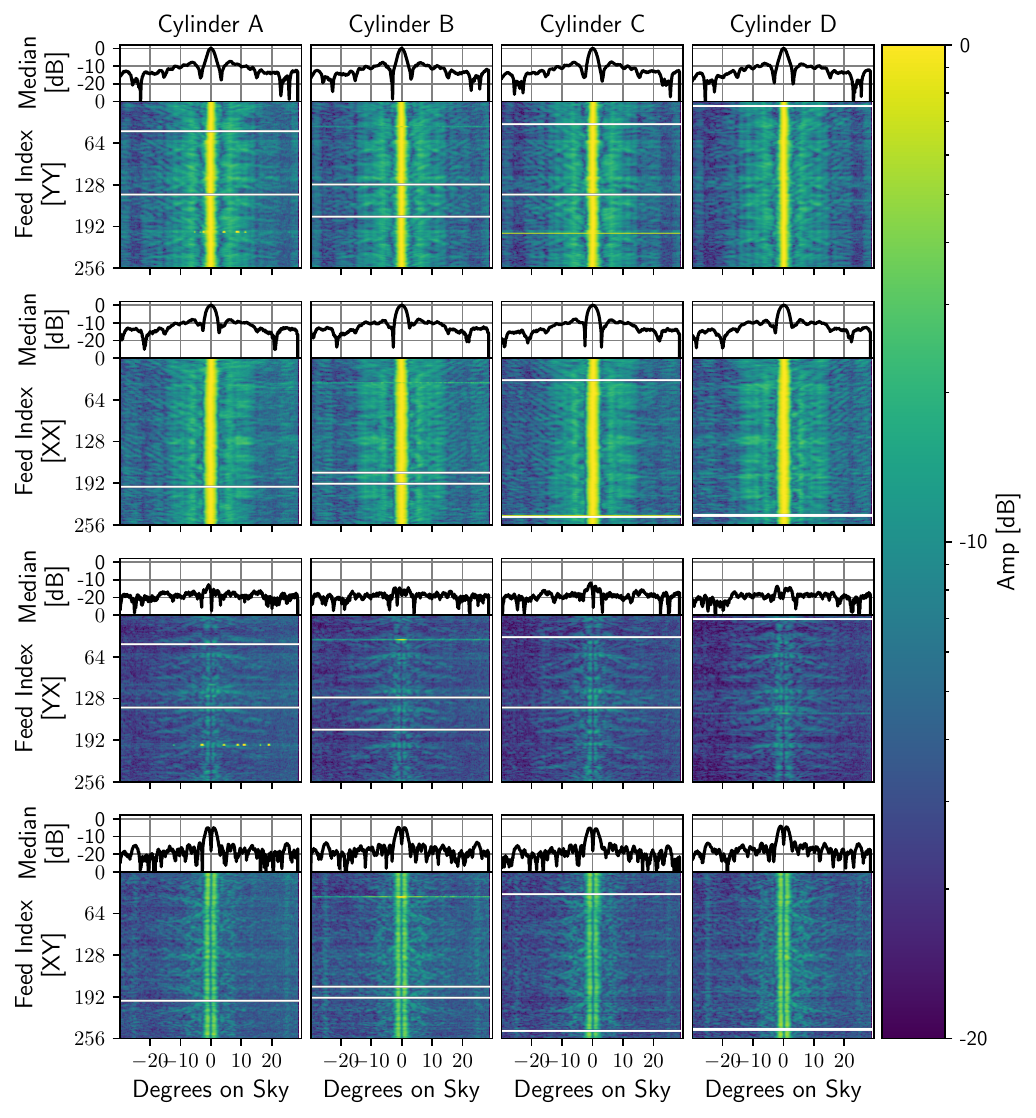}
    \caption{Waterfall plots (in feed index vs. sky degrees) of the stacked Her A data. For each cylinder data is taken from 717 MHz. The top two rows show co-pol data while the bottom two rows show cross-pol. The one dimensional slices above each image panel show the median along the respective cylinders.}
    \label{fig:feedwaterfallHer}
\end{figure*}

\begin{figure*}[ht]
    \centering
\includegraphics{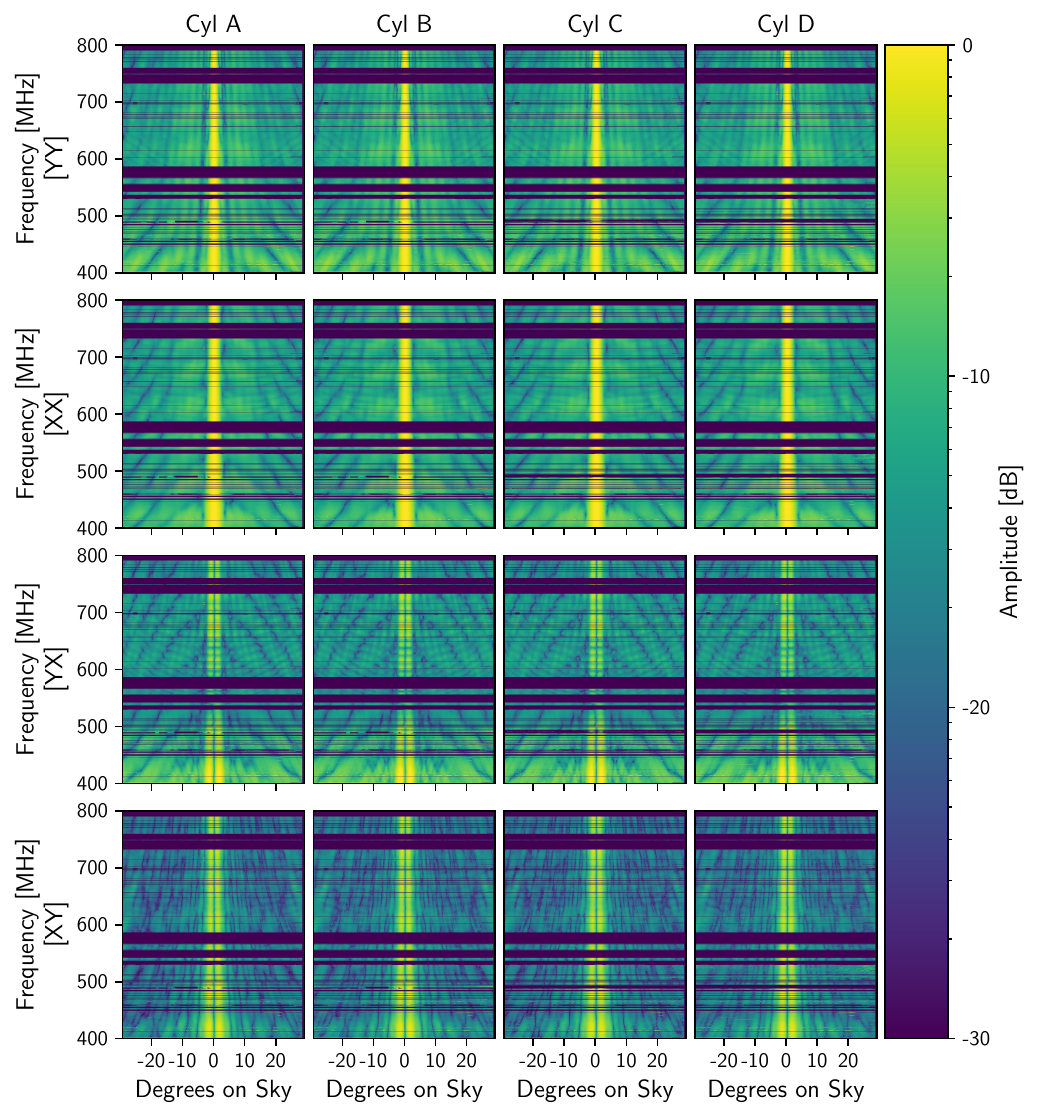}
    \caption{Waterfall plots (in frequency vs sky degrees) of the stacked Hydra A data.}
    \label{fig:freqwaterfallHyd}
\end{figure*}

\begin{figure*}[t]
    \centering
    \includegraphics{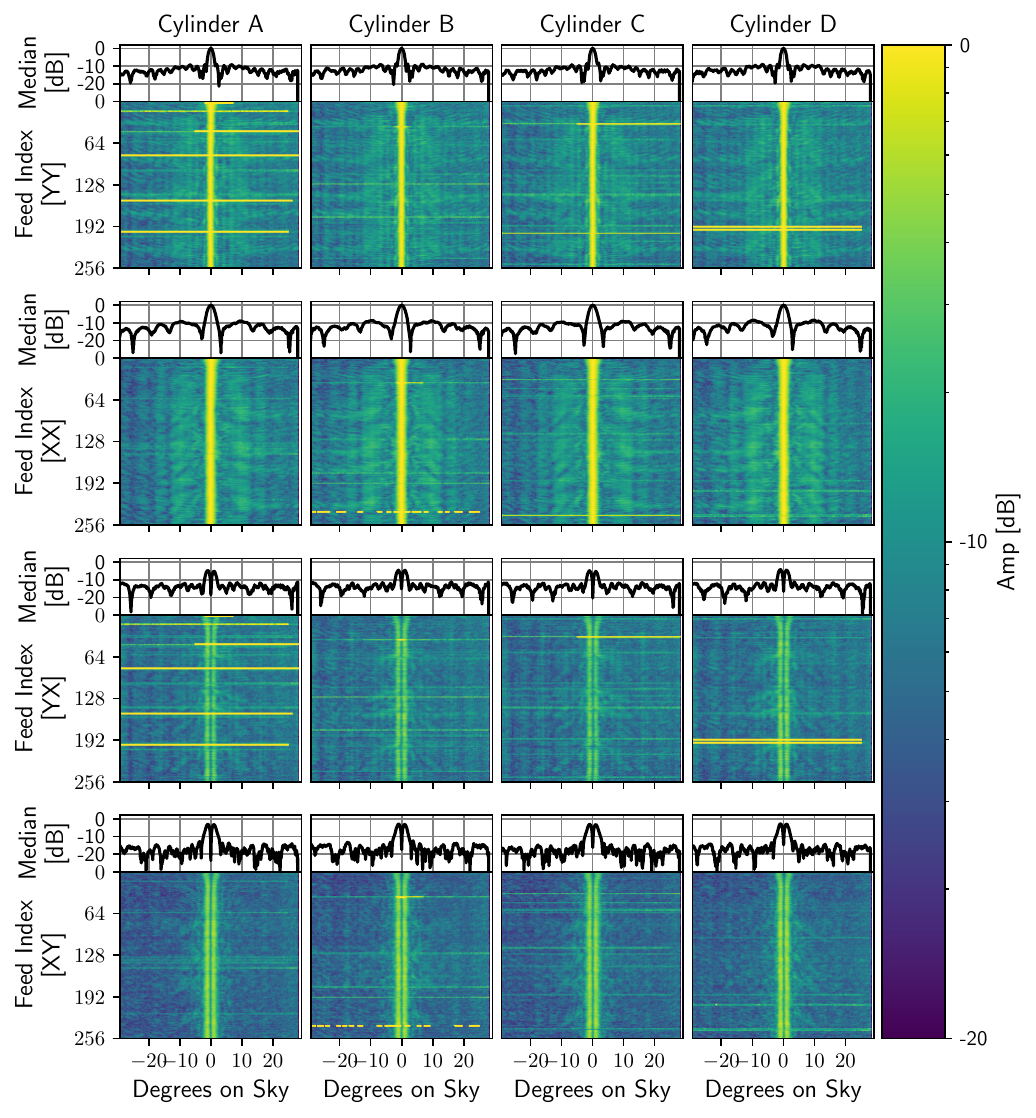}
    \caption{Waterfall plots (in feed index vs. sky degrees) of the stacked Hydra A data. For each cylinder data is taken from 717 MHz. The top two rows show co-pol data while the bottom two rows show cross-pol. The one dimensional slices above each image panel show the median along the respective cylinders.}
    \label{fig:feedwaterfallHyd}
\end{figure*}

\bibliography{main}{}
\bibliographystyle{aasjournal}

\end{document}